\documentclass[prd,singlecolumn]{revtex4}
\pdfoutput=1
\usepackage{amssymb,latexsym}
\usepackage{amsmath,amsbsy,bbm}
\usepackage{ifpdf}
\usepackage{epsfig,bm}
\usepackage{graphicx,comment}
\usepackage{color}
\usepackage{soul}
\usepackage{mathtools}
\usepackage{comment}
\usepackage[normalem]{ulem}
\begin{document}

\title{A Machine Learning study of the two-dimensional antiferromagnetic Ising model with nearest and next-to-nearest interactions on the triangular lattice }
\author{Shang-Wei Li}
\affiliation{Department of Physics, National Taiwan Normal University,
  88, Sec.4, Ting-Chou Rd., Taipei 116, Taiwan}
\author{Yuan-Heng Tseng}
\affiliation{Department of Physics, National Taiwan Normal University,
	88, Sec.4, Ting-Chou Rd., Taipei 116, Taiwan}
	\author{Kai-Wei Huang}
	\affiliation{Department of Physics, National Taiwan Normal University,
		88, Sec.4, Ting-Chou Rd., Taipei 116, Taiwan}
\author{Fu-Jiun Jiang}
\email[]{fjjiang@ntnu.edu.tw}
\affiliation{Department of Physics, National Taiwan Normal University,
88, Sec.4, Ting-Chou Rd., Taipei 116, Taiwan}

\begin{abstract}

We study the phase transitions of the two-dimensional antiferromagnetic Ising model with nearest $J_1$ and next-to-nearest $J_2$ interactions on the triangular lattice for $J_2/J_1 = 0.1, 0.5$ and 1.0. The method of supervised neural networks (NN) is employed for the investigation. While supervised NN is used, no real spin configurations are needed for the training. In addition, two
kinds of configurations having their spins be arranged in a staggered pattern are considered as the training set. Remarkably, with this unconventional training strategy, not only the critical temperatures of the studied $J_2/J_1$ are computed accurately by the resulting NN, but also the nature of the investigated phase transitions are determined correctly. Specifically, the phase transitions associated with $J_2/J_1 = 0.1, 0.5$ and 1.0 are first order. These conclusions are consistent
with the known results obtained by other methods. Since the training
strategy is simple, the NN calculations is highly efficient. It remains to examine whether the unconventional training
approach considered in this study can be used to investigate other models with untypical phase transitions or with nontrivial ground state configurations.

\end{abstract}

\maketitle

\section{Introduction}

Ising model is one of the most studied spin systems in physics \cite{Ons44,McC14,Xu18}.
This is largely due to the simplicity of this model. Hence, one can
compute the associated properties, such as the critical temperature $T_c$ and the critical exponents of the Ising model with high accuracy using either analytic or numerical methods. In conclusion, This model is among the well-understood models. 

Ising model has been considered in various spatial dimensions and lattice geometries. In addition to the interactions between
any pair of nearest neighboring Ising spins (can be either ferromagnetic or antiferromagnetic ones), the introduction of
couplings between two next-to-nearest neighboring spins and
terms of higher order interactions (can be either ferromagnetic or antiferromagnetic ones as well) make the characteristics of these variant of Ising models, specially their phase transitions and ground state configurations, intriguing \cite{Met73,Tab75,Bin76,Bin80,Lan80,Oit81,Lan83,Slo84,Lan85,Mer04,Ras05,Mal07,Kal08}. Besides, they may be relevance to describe real systems such as the adsorbate phases of crystals as well \cite{Bin76,Bin80}.

Recently, techniques of Machine Learning has demonstrated their
power in exploring the physics of many-body systems. It has been
proved that the neural networks (NN) methods can be considered as
an alternative to compute the features of critical phenomena, such as the associated critical points, for many physical models \cite{Tan16,Car16,Nie16,Meh19,Car19,Cor21,Kim18,Li18,Tan20,Tan20.1}. 

Conventionally, the application of a NN to study a phase transition
consists of 3 stages, namely the training, the validation, and the
testing stages. We refer readers to Refs.~\cite{Meh19,Car19} for the detailed descriptions of these stages. 

Among the three stages mentioned in the previous paragraph, 
the training stage is the most time-consuming one. First, a NN architecture should be designed. Next, real configurations
of the studied system (These are called the training set), which may be a huge file,
are used to train the built NN with sufficiently large number of
epochs. This is to ensure the features of different phases are 
captured well by the NN. The training set should also contains the configurations from both sides of the critical point. Based on the experiences of the authors, the training procedure may
take several hours to few days. The required time for training a NN depends on how many 
configurations are included in the training set as well as the complexity of the NN architecture. Intuitively, one expects that for a NN to perform well in exploring the critical phenomena, the NN should have a complex architecture and the training set used ought to contain large number of configurations. 

Interestingly, it has been demonstrated that a simple multilayer perceptron (MLP), which consists of one input layer, one hidden 
layer of two neurons (or other number of neurons), and one output layer, and is trained with two kinds of artificially made (ferromagnet-type) configurations, can successfully
detect the critical points and compute the associated critical exponents with good accuracy for many two-dimensional (2D) and three-dimensional (3D) models having ferromagnet-like interactions \cite{Tan20.1,Tse22, Tse23, Pen23,Tse241,Tse242,Jia24}. 
It is really remarkable that by elegantly constructing the testing set
from the original raw spin configurations, a single simple MLP can 
compute the critical properties of many models with ferromagnet-type interactions differing among themselves dramatically.

The mentioned universal MLP has been employed to investigate the critical behaviors of 2D antiferromagnetic 2-state, 3-state, and 4-state Potts models on the square lattice \cite{Tse25}, and the NN outcomes
are consistent with the expected theoretical predictions. In other words, it also works for models for which the related interactions between any pairs of nearest spins are not of ferromagnet-type. 

Still, it would be worthy the effort to examine whether a NN trained with other kinds of artificially made configurations can be considered to uncover the critical phenomena of models having
antiferromagnetic ground states or more complicated ground states such
as the one with stripe pattern. The stripe configurations are the ground states
of the 2D frustrated Ising model with ferromagnetic nearest coupling $J_1$ and antiferromagnetic next-to-nearest coupling $J_2$ on the square lattice as well as the 2D Ising model which has both antiferromagnetic $J_1$ and $J_2$ on the triangular lattice (The latter
model is called the 2D antiferromagnetic $J_1$-$J_2$ model on
the triangular lattice in this study for convenience).   

Here, we train a MLP having one input layer, one hidden layer, and one output layer using two types of artificially made configurations
as the training set. In particular, the values of spins for the configurations in the training set take the values of 1 and -1 in a staggered pattern. The obtained MLP is then applied to study the phase transitions of 2D Ising model which has both antiferromagnetic $J_1$ and $J_2$ on the triangular lattice with $J_2/J_1 = 0.1, 0.5$ and 1.0. 

Remarkably, the MLP not only detects the phase transitions for all the three considered values of $J_2/J_1$, but also determines the nature of these phase transitions to be first order without ambiguity.
This provides evidence to the potential applicability of this unconventionally trained MLP to study models with non-trivial ground states. To verify this, it remains to investigate the phases transitions of the 2D antiferromagnetic 3-state and 4-state Potts models, and the 2D frustrated Ising models with ferromagnetic nearest coupling $J_1$ and antiferromagnetic next-to-nearest coupling $J_2$ on both the square and the honeycomb lattices.

Finally, it should be pointed out that when the traditional methods are considered to study phase transitions, the relevant order parameters should be known in advance before one can carry out the investigations. This is not the case for the NN approach. In other words,
one can conduct the study of critical phenomenon of a phase transition with NN without any information of the associated order parameter. This will be elaborated later in the section " Discussions and Conclusions ".

The rest of the paper is organized as follows. After the introduction, the antiferromagnetic $J_1$-$J_2$ Ising model on the triangular lattice,
the used MLP, the training details, and the construction of testing sets are described in Sects. II and III, respectively.
Then in Sect. IV, the numerical outcomes are presented. Finally, we conclude our study in Sect. V.

\section{The considered model}

\begin{figure}
    \includegraphics[width=0.5\textwidth]{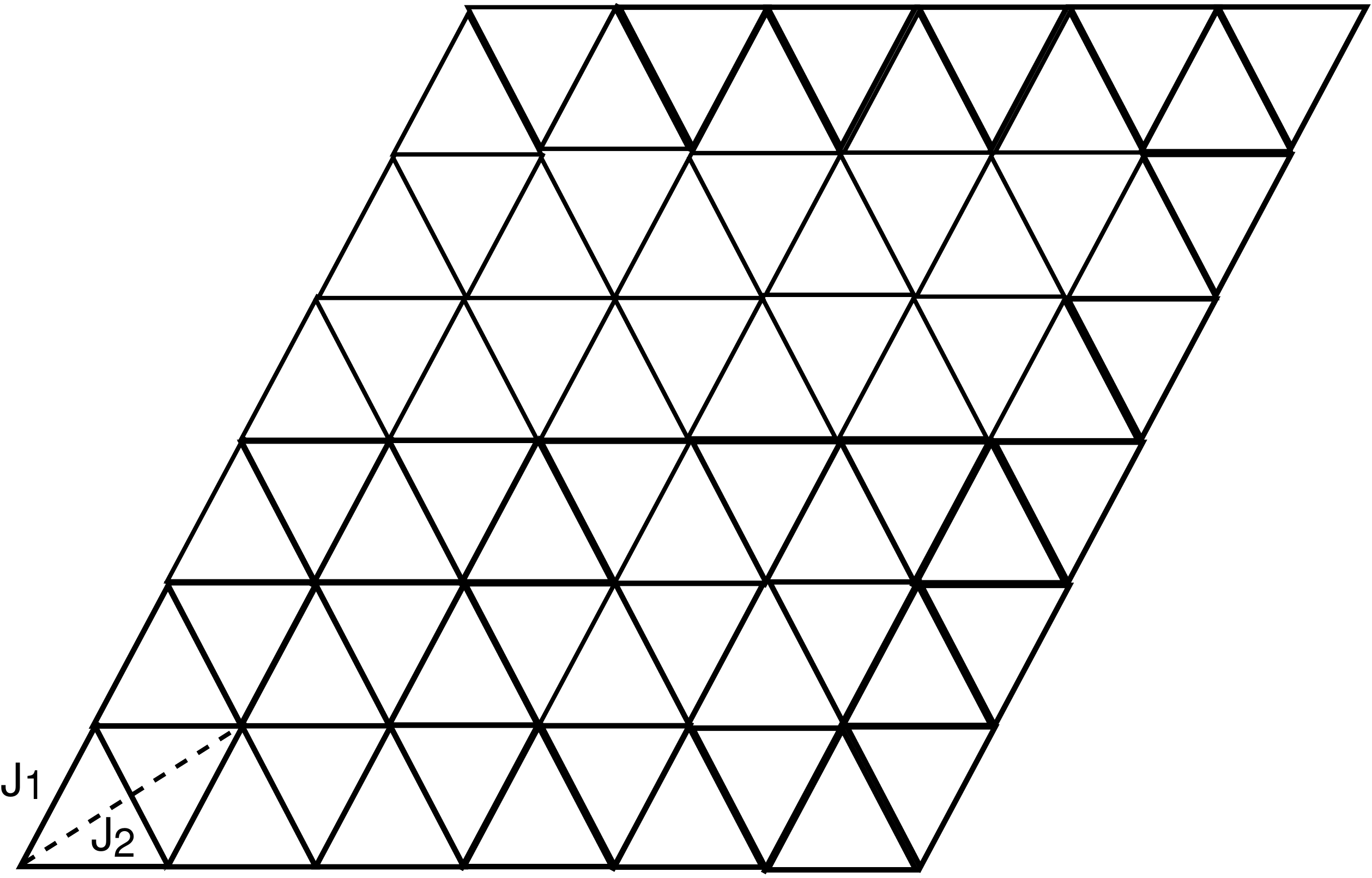}
        \caption{The antiferromagnetic $J_1$-$J_2$ Ising model on the triangular lattice studied here. The solid and the dashed lines represent the nearest coupling $J_1$ and next-to nearest coupling $J_2$, respectively. 
        Only one $J_2$ is shown. }
        \label{j1j2}
\end{figure}

The Hamiltonian $H$ of the antiferromagnetic $J_1$-$J_2$ Ising model with nearest coupling $J_1$ and next-to-nearest coupling $J_2$ on the triangular lattice considered in this study takes the form
\begin{equation}
 H = J_1 \sum_{\left< ij\right>} \sigma_i\sigma_j + J_2 \sum_{\left<\left< lm\right>\right>}\sigma_l\sigma_m,
\label{eqn}
\end{equation}
where $\sigma_i = \pm 1$ is the Ising spin at site $i$. The first
and the second terms appearing in Eq.~(\ref{eqn}) are summed over nearest neighboring sites $i$ and $j$ and summed over next-to-nearest neighboring sites $l$ and $m$, respectively.
Fig.~\ref{j1j2} is the graphical representation of the model studied here. It should be pointed out that in this study a triangular lattice consisting of $L_1 \times L_2$ sites ($2 \times L_1 \times L_2$ triangles) is embedded in a $L_1$ by $L_2$ parallelogram area,
and periodic boundary conditions are used. For example, the triangle lattices in fig.~\ref{j1j2} has 36 sites (72 triangles) and they are embedded in a 6 by 6 parallelogram. Here both $L_1$ and $L_2$ are multipliers of 6.

In our calculations, $J_1 = 1$ and $J_2 > 0 $. The phase
transitions of $J_2 = 0.1$, $0.5$ and $1.0$ are investigated.
It is established that the $T_c$ for $J_2 = 0.1$, $0.5$, and 1.0 are given by 0.5756, 1.374 and 1.808, respectively \cite{Ras05}. Moreover, all these phase transitions are first order. In particular, while $J_2 = 0.5$ is a weak first-order phase transition, both the phase transitions of $J_2 = 0.1$ and $J_2 = 1.0$ show strong signal of being first order.

\section{The employed NN}

\begin{figure}
	\includegraphics[width=0.8\textwidth]{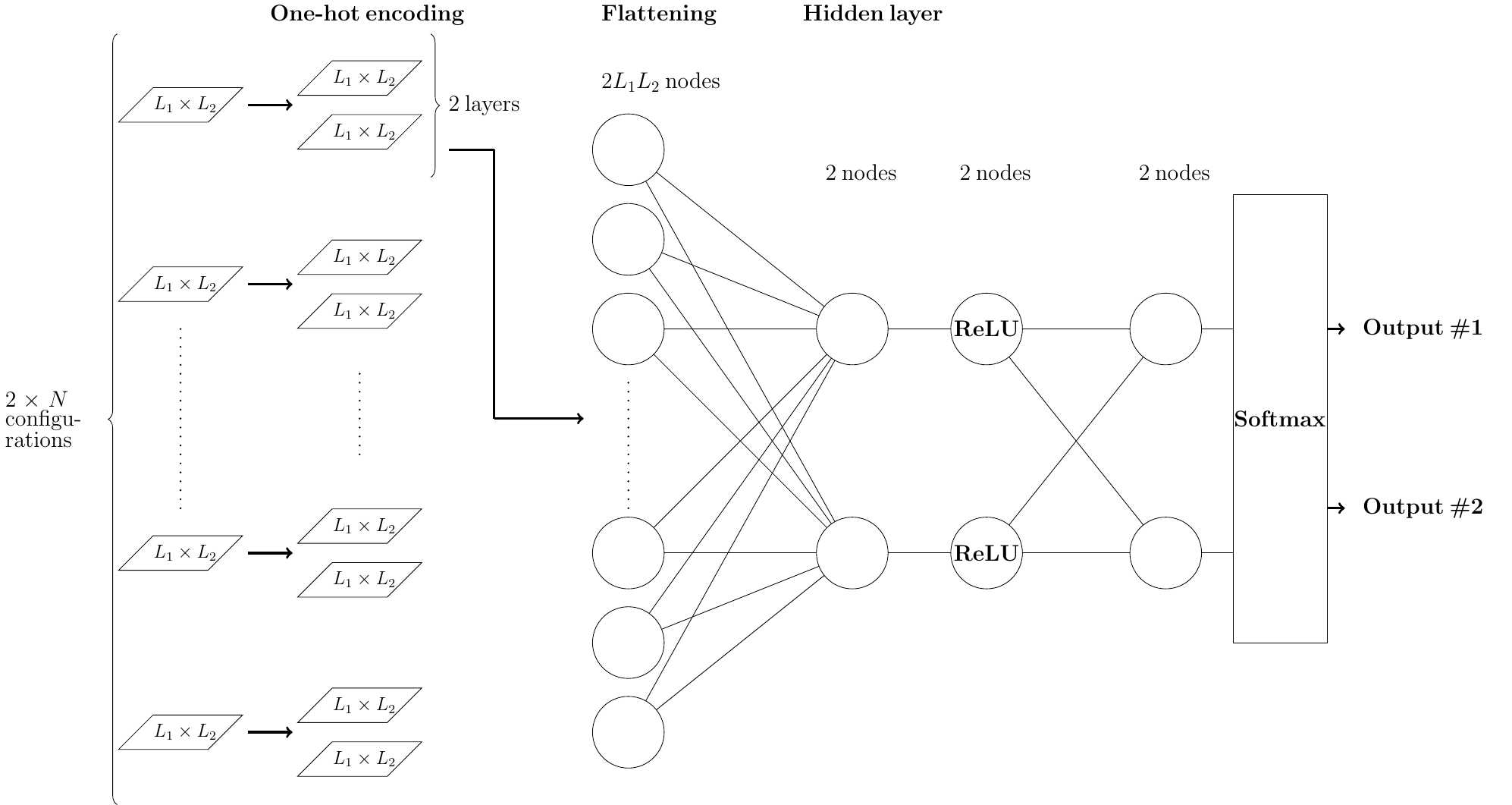}
	\caption{Representation for one of the MLPs used in this study. $L_1$ and $L_2$ are the linear system sizes for the parallelogram covering a triangular lattice having $L_1 \times L_2$ sites. $N$ can be any number and here we use $N=200$ or $N=400$. This figure is drawn based on the MLP of Ref.~\cite{Tan20.1}.}
	\label{nn}
\end{figure}

The NN considered here consists of one input layer, one hidden layer of two neurons (or other number of neurons), and one output layer. Fig.~\ref{nn} is the pictorial
representation of one of the employed NNs. In particular, the activations functions for the hidden layer and the ouput layer are
ReLU and Softmax, respectively. Before data enter the hidden layer, the procedure of one-hot encoding is used. To prevent overfitting, in the dense (hidden) layer, $L_2$ regularization for the kernel with the input parameter being 1 is considered as well. The algorithm and optimizer employed are minibatch and adam, respectively. Finally,
the MLP outputs are two-component vectors.

\subsection{Training procedure}

\begin{figure}
	\includegraphics[width=0.6\textwidth]{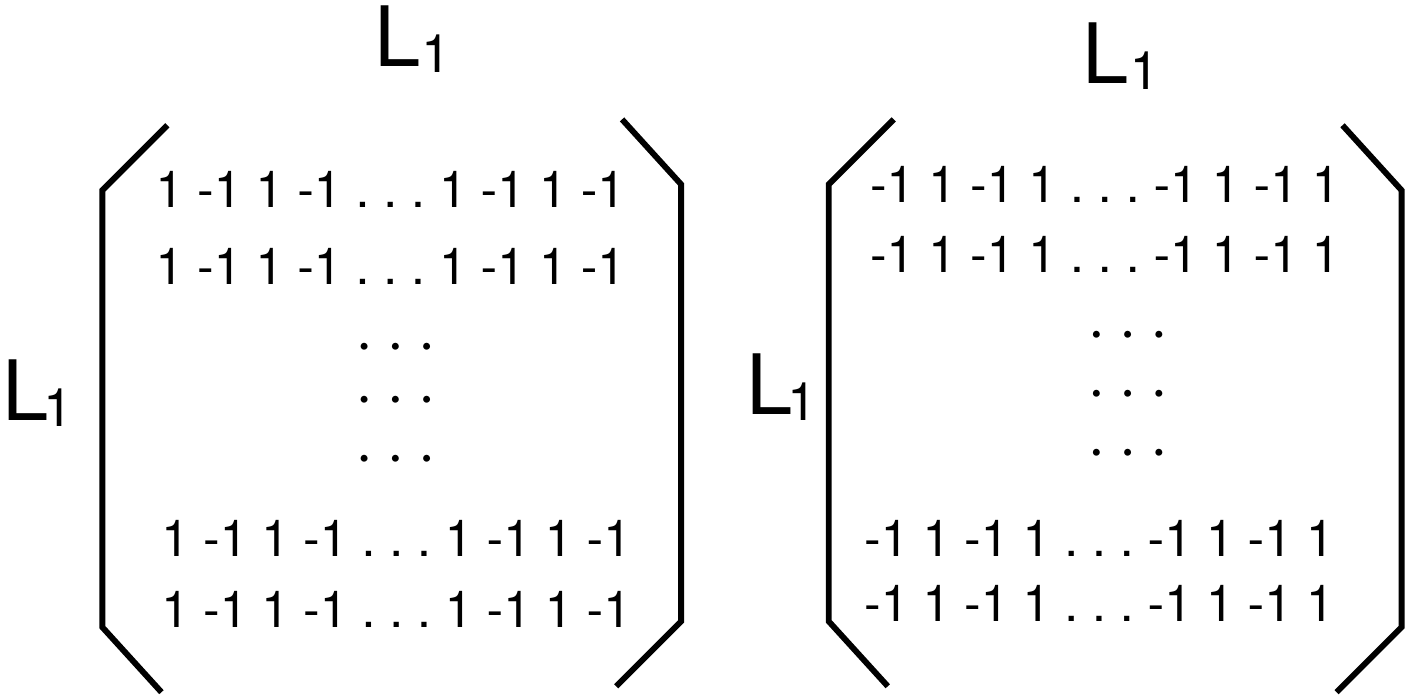}
	\caption{The building blocks of the training set employed in this study. The value of $L_1$ can be varied. }
	\label{train}
\end{figure}

Instead of using real configurations generated from Monte Carlo simulations, here we consider two kinds of artificially made configurations as the training set. Fig.~\ref{train} shows two building block of the training set used here. Specifically,
the training set consist of $N$ ($N=200$ or 400 in this study) copies of both the left and the right panels of fig.~\ref{train}. Here each building block is made up of $L_1$ by $L_1$ sites and the value of $L_1$ can be varied. In addition, the MLP output labels for the left and the right panels are $(1,0)$ and $(0,1)$, respectively. Finally, we use at least 800 epochs for the trainings. 

\subsection{The construction of testing set}

\begin{figure}
	\includegraphics[width=0.5\textwidth]{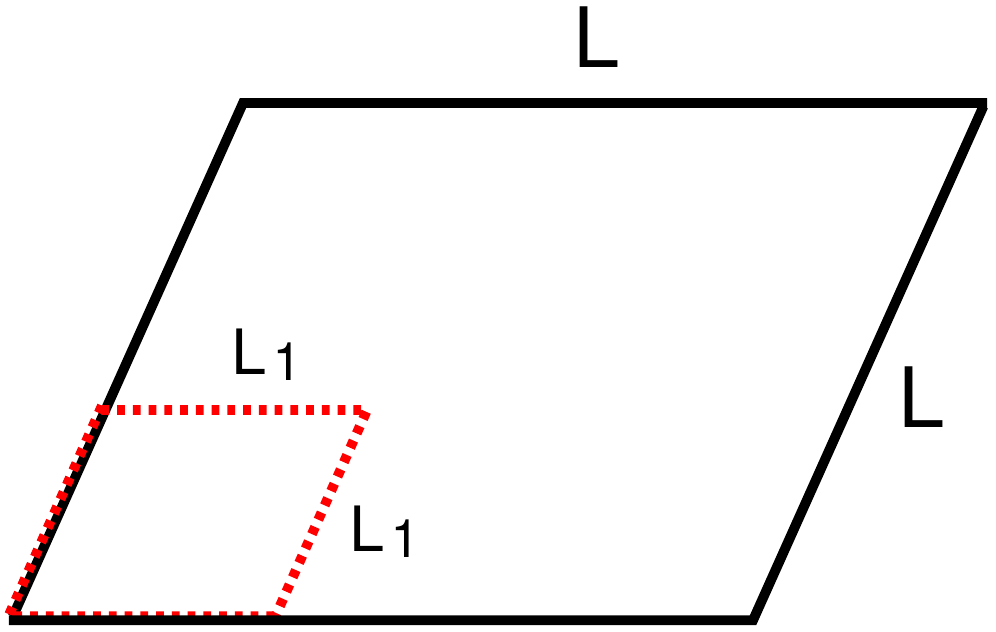}
	\caption{The construction of a configuration for the testing set related to a given linear box size $L_1$ from a spin configuration on a $L$ by $L$ parallelogram. }
	\label{testing}
\end{figure}

For a configurations on a $L$ by $L$ parallelogram generated from the Monte Carlo simulations, the spins belonging to the first $L_1$ by $L_1$ sub-parallelogram are considered as one of the testing set for the given $L_1$. The representations of the original configurations (on a $L$ by $L$ parallelogram) and the resulting $L_1$ by $L_1$ configuration used as one of the testing set are depicted in fig.~\ref{testing}. Notice while $L$ is a multiplier of 6,
we put no such a restriction on $L_1$. 

For convenience, when the 
NN outcomes are presented in the following, $L$ and $L_1$ will be 
used to stand the linear system sizes of the configurations generated by the Monte Carlo calculations and the configurations of the testing
(and the training) set, respectively.

\subsection{The magnitude $R$ of the MLP output vectors}

For a vector $\vec{A} = (A_1,A_2)$, let the associated magnitude be
denoted by $R$, i.e. $R = \sqrt{A_1^2 + A_2^2}$.
In the testing stage, if the input configuration is highly similar to the training set, then the output vector is close to $(1,0)$ or $(0,1)$ which would result in $R \sim 1$. On the other hand, if the
input configuration is in strong contrast to the training set, then
the output vector is close to $(0.5,0.5)$ which would lead to $R \sim 1/\sqrt{2}$. In other words, by investigating how $R$ changes with $T$, one can estimate the value of the critical point. For a detailed explanation of how this works, see Refs.~\cite{Tan20.1,Tse22,Tse23,Tse241,Tse242,Jia24,Tse25}.  

\section{Numerical Results}

To study the proposed phase transitions, for each of $J_2 = 0.1, 0.5$, and 1.0, and every given triangular lattice having $L$ by $L$ spins, we have generated few to several thousand configurations using single spin flip Metropolis algorithm \cite{Met53} at various temperatures. Moreover, we would like to emphasize the fact that for temperatures far below the critical temperature $T_c$, although one of the equilibrium steady ground states can be obtained by the MC simulations, sampling among these ground states is difficult because the associated MC chains are generated by a local update algorithm. 

\begin{figure}                                               \vbox{             
		\hbox{~~~~~~~~
	\includegraphics[width=0.4\textwidth]{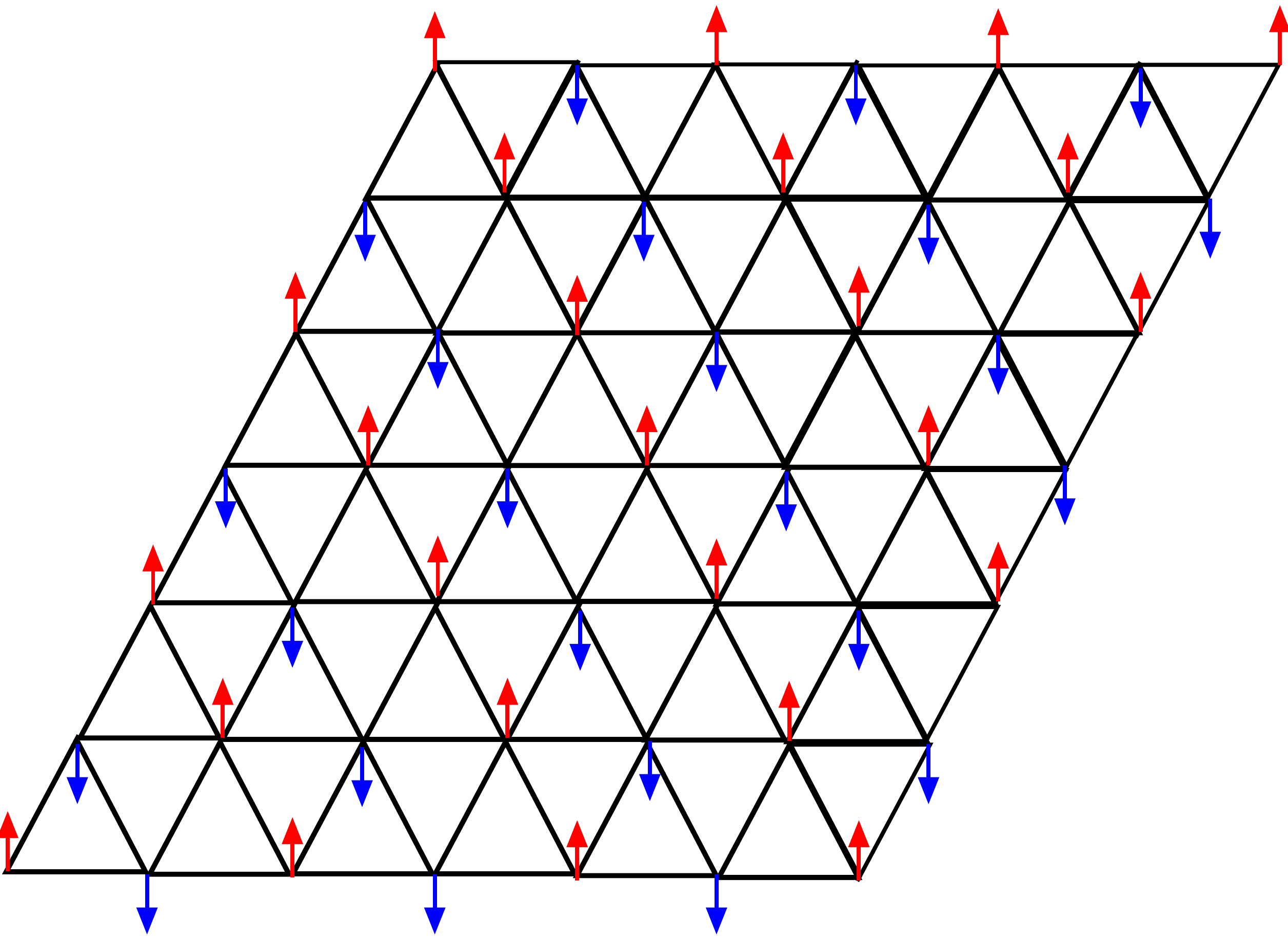}~~~~~
	\includegraphics[width=0.4\textwidth]{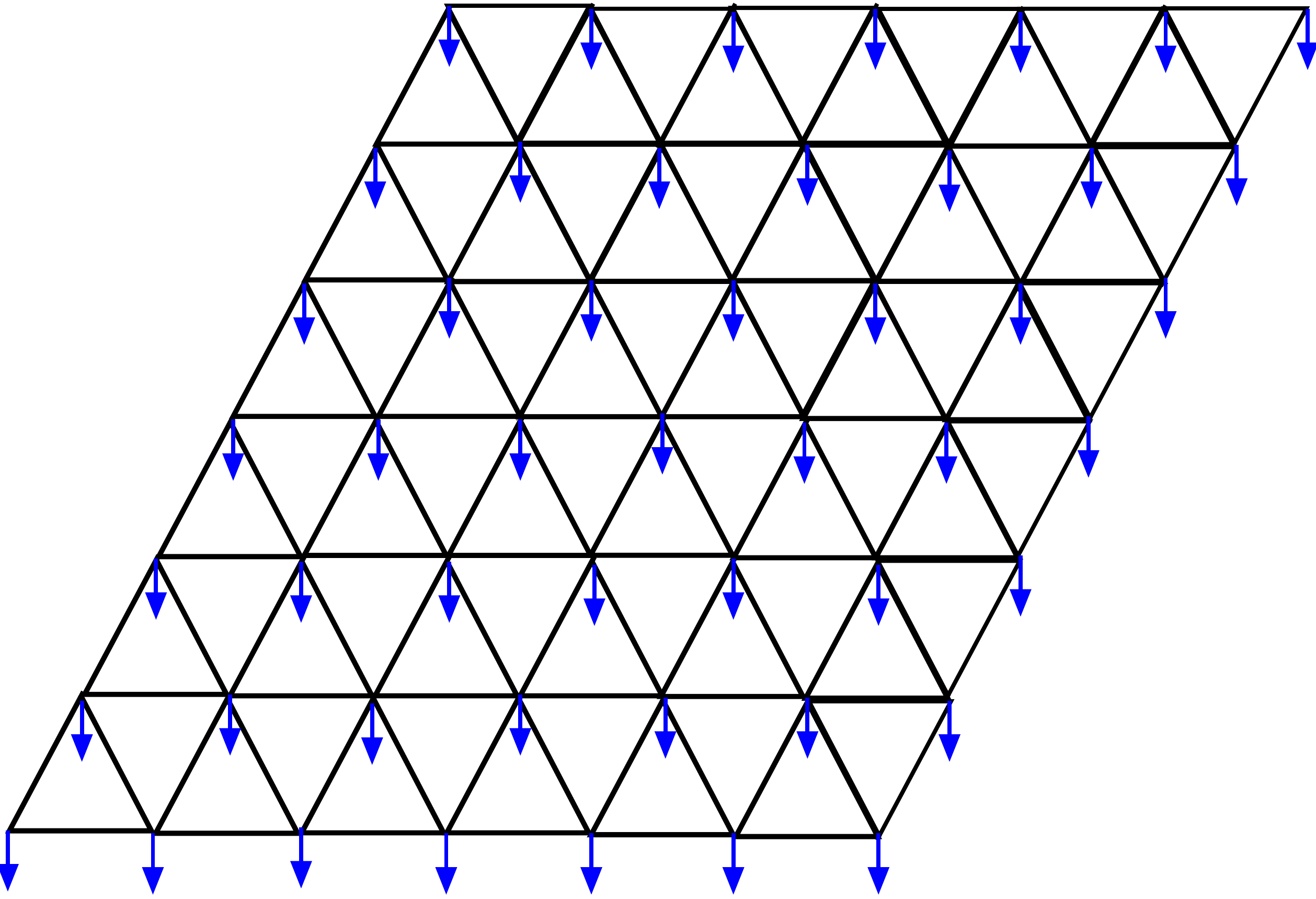}}
	\hbox{~~~~~~~~
	\includegraphics[width=0.4\textwidth]{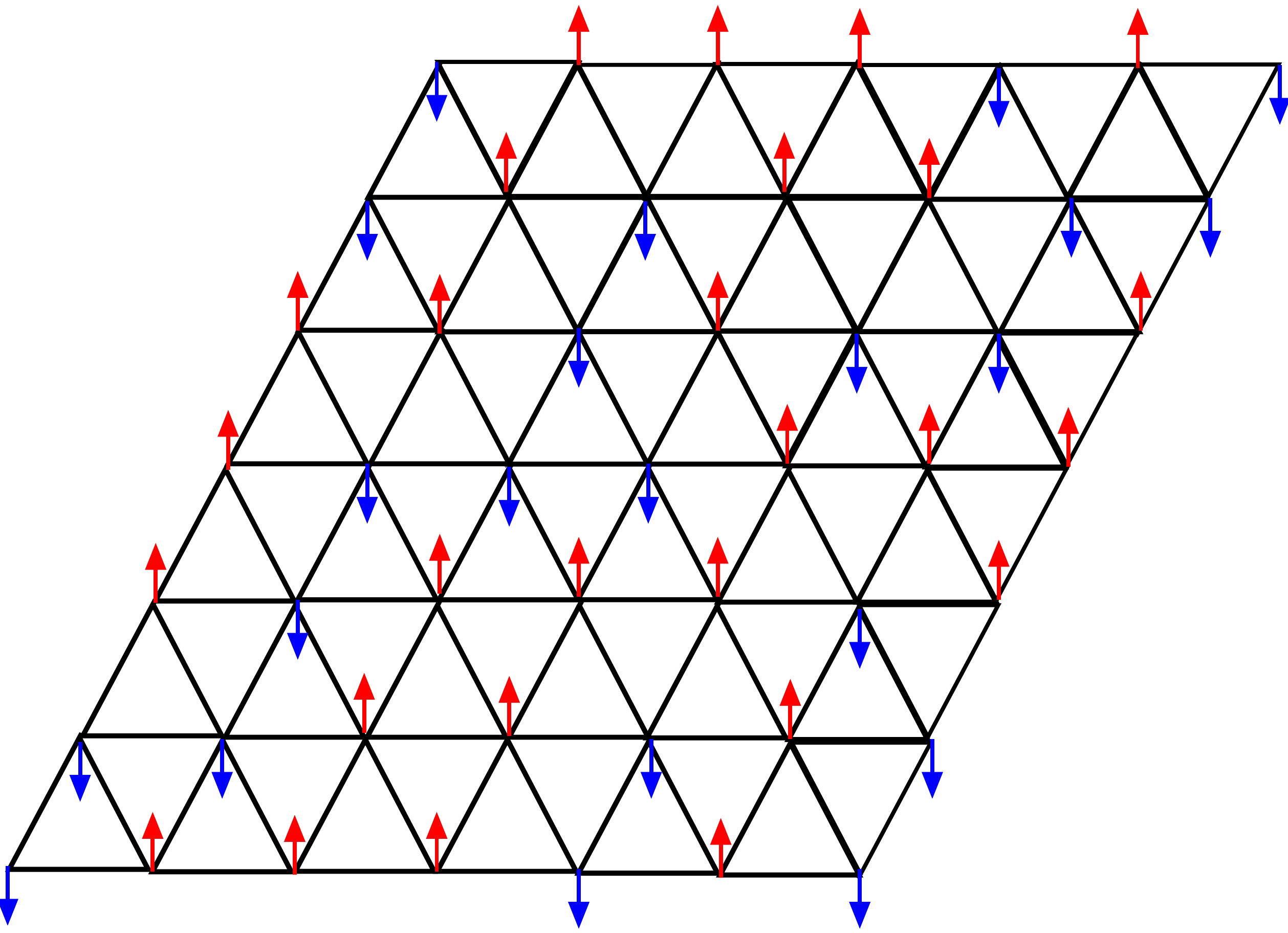}~~~~~
\includegraphics[width=0.4\textwidth]{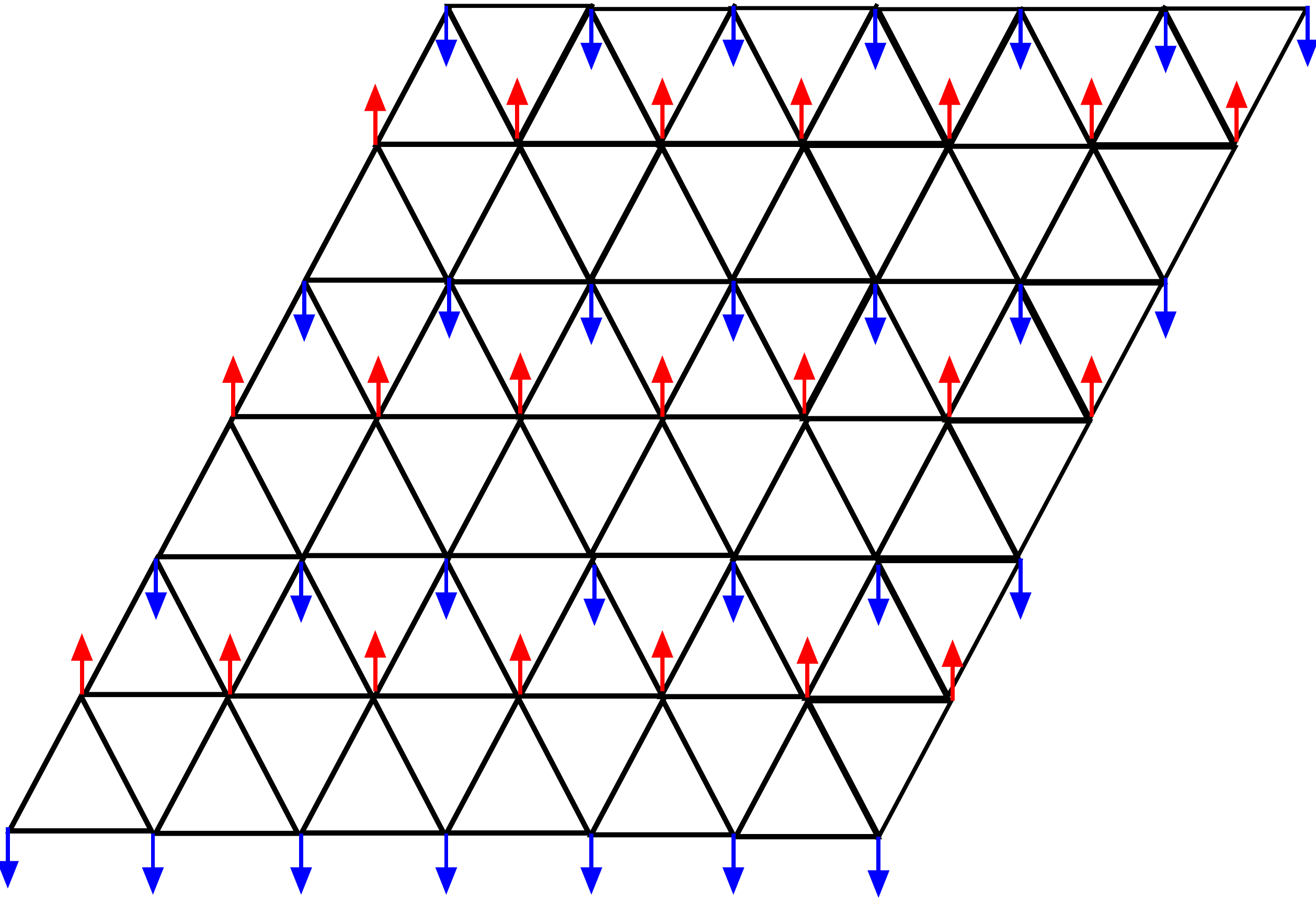}
}
}
	\caption{The initial configurations (on a 6 by 6 parallelogram), namely stagger-start (top left panel), uniform-start (top right panel), and randomness-start (down left panel) used to begin the MC simulations. Arrows which point upward (downward) stand for spins with their values being 1 (-1).}
	\label{initial}
\end{figure}

Despite the fact mentioned in the previous paragraph, as we will demonstrate shortly, the inclusion of the configurations at low-$T$
region is very useful in determining the true ground states and estimating the critical temperatures. Finally, due to the inefficiency of the algorithm, three-types of configurations are used as the initial configurations to start the Monte Carlo simulations,
see fig.~\ref{initial} for these 3-types of configurations and for convenience, they are called the stagger-, the randomness-, and the uniform-start, respectively. The names for these initial configurations can be comprehended with ease from the associated
cartoon representations shown in fig.~\ref{initial}. For example,
all the spins point downward in the configuration is named 
uniform-start. Moreover, for the stagger-start configuration, the  values of the related spins are alternative in 1 and -1 along both the rows and the columns.

Before demonstrating our NN results, we would like to point out that 
 based on the facts that the triangular lattices are embedded in parallelogram areas and how the periodic boundary conditions are implemented here, the spins of the ground states for the considered model(s) have either a staggered pattern (The top left panel of fig.~\ref{initial}) or a stripe pattern (The down right panel of fig.~\ref{initial} is one of the four configurations having the stripe pattern). This result associated with the theoretical ground state configurations is different from that of Ref.~\cite{Ras05}. Later we will make some comments on this.  
 
\subsection{$T$-dependence of $R$ for first-order and second-order phase transitions}

\begin{figure}
    \hbox{             
	\includegraphics[width=0.33\textwidth]{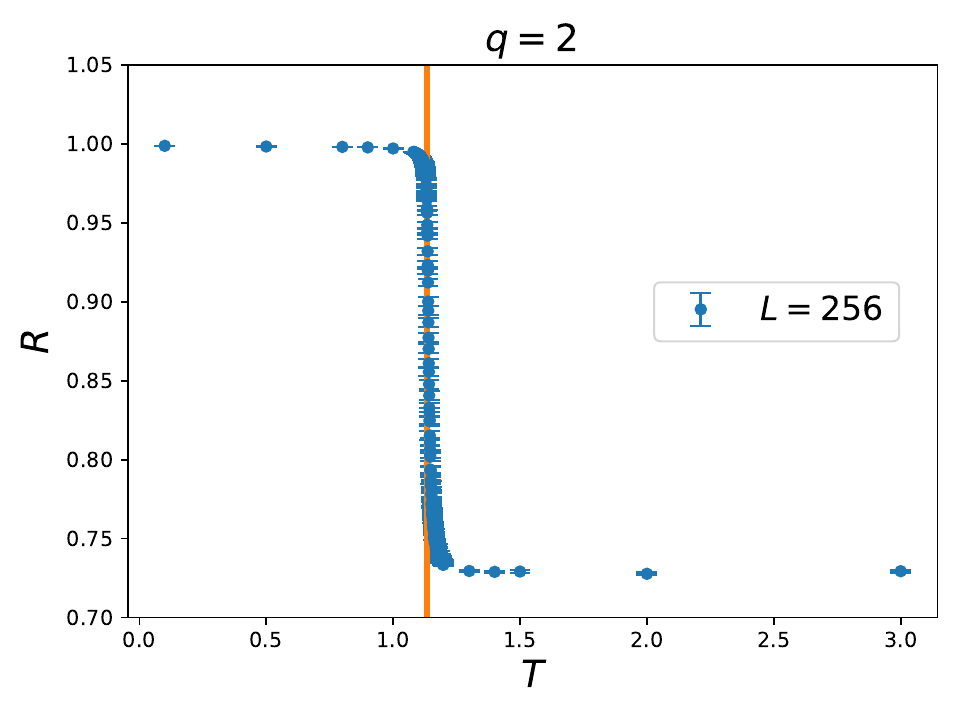}
	\includegraphics[width=0.33\textwidth]{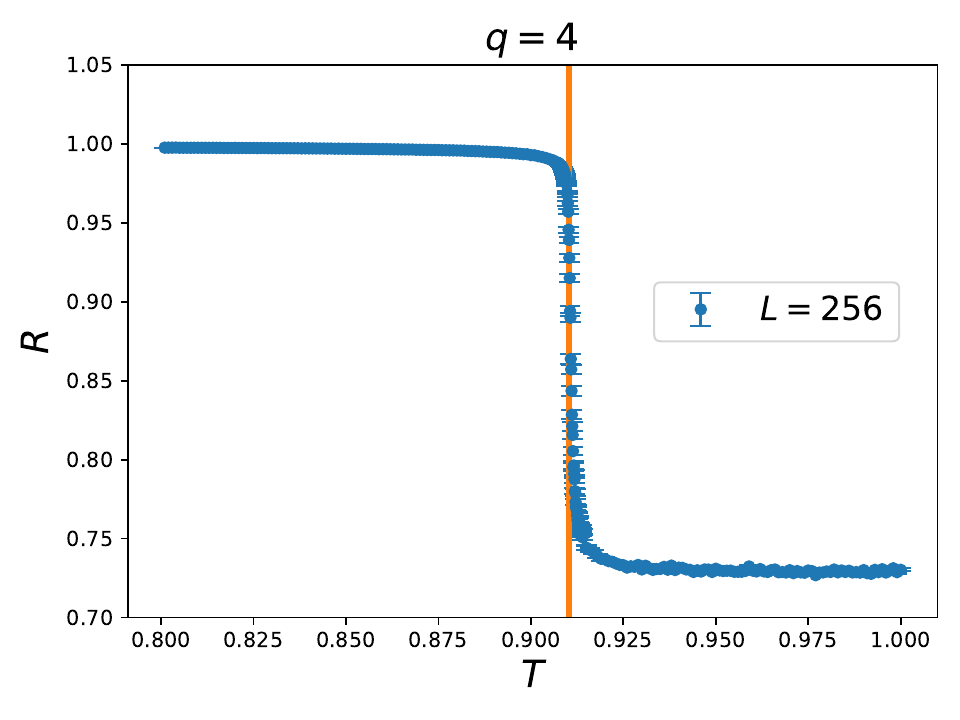}
	\includegraphics[width=0.33\textwidth]{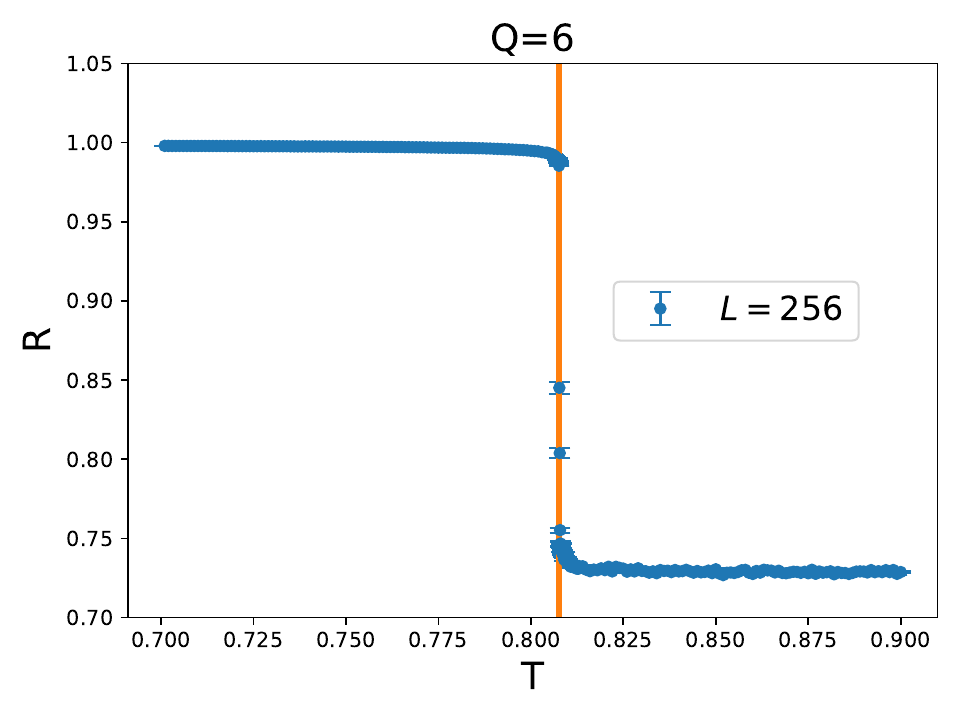}	
}
	\caption{$R$ as functions of $T$ for the two-state (left), the four-state (middle), and the six-state ferromagnetic Potts models on the square lattice. The vertical solid lines are the expected values of $T_c$. These figures are taken from Ref.~\cite{Tse241}.}
	\label{potts}
\end{figure}

The left, the middle, and the right panels of fig.~\ref{potts} show the $T$-dependence of
$R$ for the 2D two-state ($q=2$), four-state ($q=4$), and six-state ($q=6$) ferromagnetic Potts models on the square lattice (These figures are taken from Ref.~\cite{Tse241}). The vertical solid lines in these panels are the expected values of $T_c$. Theoretically, it is well-known that the phase transitions of $q=2$ and $q=4$ models are second order, and the phase transition of $q=6$ model is first order. As can be seen in fig.~\ref{potts}, the $T$-dependence of $R$ not only computes the $T_c$ accurately, but also faithfully reveals the nature of the phase transitions for these models. Indeed, the values of $T$ where rapid changes in $R$ match well with the expected $T_c$ excellently. Moreover, close to $T_c$,
while $R$ varies continuously with $T$ for $q=2$ and $q=4$ models,  
it has a sudden jump for $q=6$ model. These behaviors of $R$ near $T_c$ are exactly the features of second-order and first-order phase transitions. To conclude, investigations of the behaviors of $R$ close to $T_c$ can be used to compute the $T_c$ precisely and to determine the nature of the phase transitions correctly.

We would like to emphasize the fact that the results shown in fig.~\ref{potts} are obtained by a similar MLP as that shown in fig.~\ref{nn}. Moreover, the associated training set is like the one used here. In particular, the training set consists of artificially constructed ferromagnet-type configurations having two variables 0 and 1. We refer the readers who are interested in the details to Refs.~\cite{Tan20.1,Tse22,Tse23,Tse241,Tse242,Jia24}.
    
\subsection{The NN outcomes related to $J_2=0.1$}

\begin{figure}
	\hbox{             
		\includegraphics[width=0.33\textwidth]{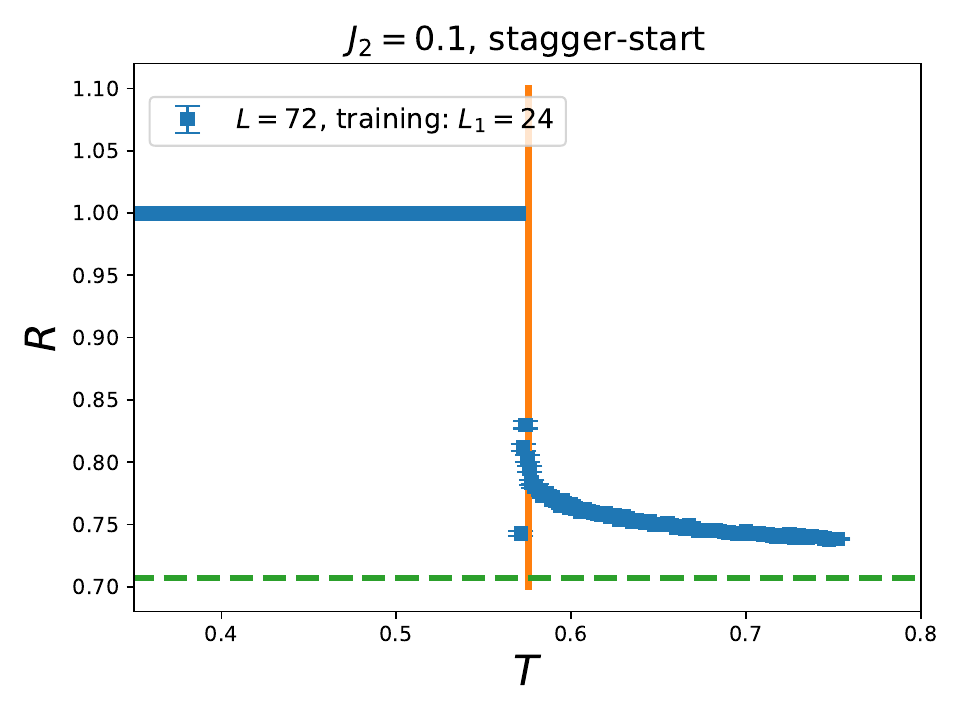}
		\includegraphics[width=0.33\textwidth]{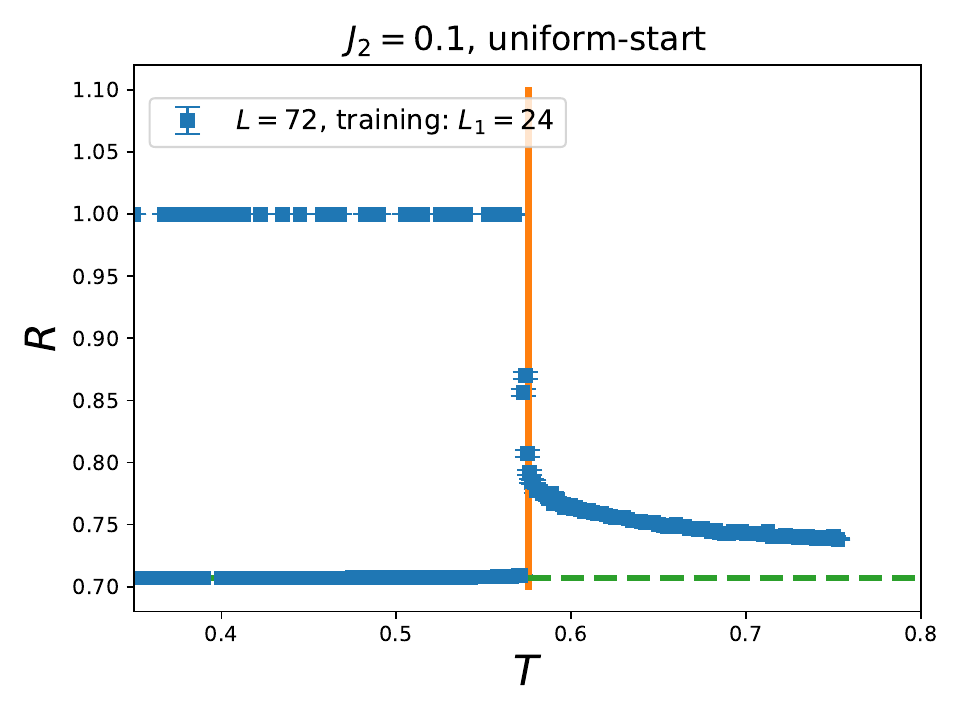}
		\includegraphics[width=0.33\textwidth]{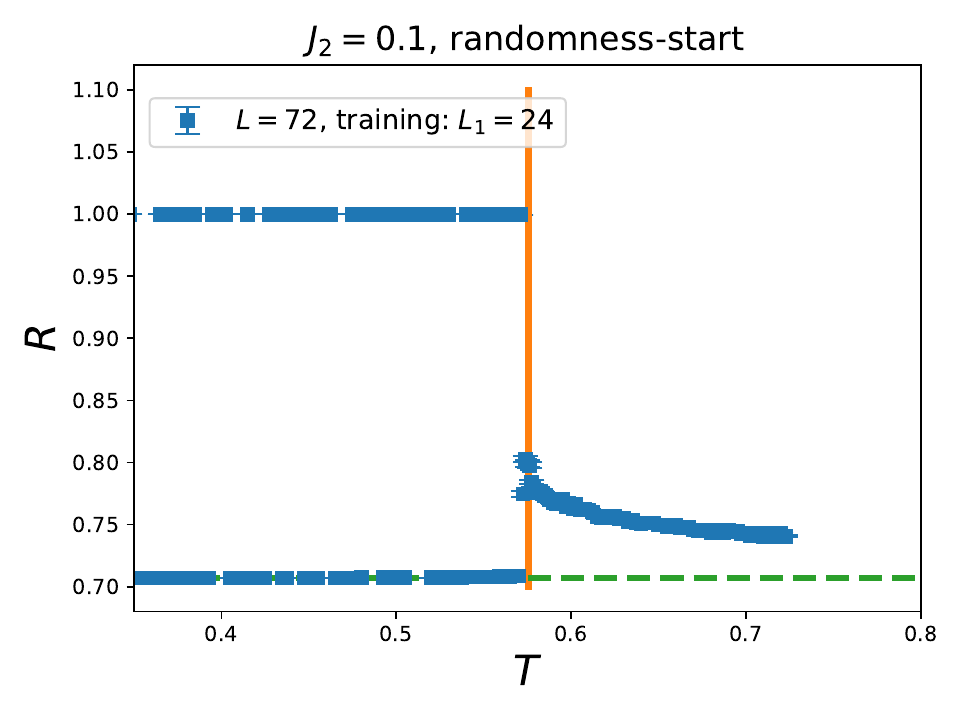}	
	}
	\caption{$R$ as functions of $T$ for the $g=0.1$, $L=72$, and $L_1=24$. The vertical solid lines are the expected $T_c$. The dashed horizontal lines are $1/\sqrt{2}$ which is the possible smallest value of $R$.}
	\label{g01L72L24}
\end{figure}

$R$ as functions of $T$ for $L=72$ and $L_1=24$ are shown in
fig.~\ref{g01L72L24}. In addition, the initial configurations used to begin
the Monte Carlo simulations are listed in the titles of the three
panels of fig.~\ref{g01L72L24}. The vertical solid lines in fig.~\ref{g01L72L24} are the $T_c$ for $g=0.1$ determined in Ref.~\cite{Ras05}. Clearly, sudden jumps in $R$ take place at a value of $T$ that agrees well with the expected $T_c$. This indicates the transition is first order beyond any doubt. The dashed horizontal lines are $1/\sqrt{2}$ which is the possible smallest value of $R$.

The results presented in
all three panels of fig.~\ref{g01L72L24} provide convincing evidence
to support the facts that our MLP, trained without using any real configurations of the considered models, can detect the $T_c$ precisely and can determine the nature of the phase transition correctly.

It should be pointed out that for $T < T_c$, the MLP outcomes $R$ associated with the
stagger-start (the left panel of fig.~\ref{g01L72L24}) always take
the value of 1. This can be understood as follows. The initial
configuration used to begin the Monte Carlo simulations is one of
the ground states of the studied model. Since the algorithm employed
can perform only local update and the probability of making
change (i.e. flipping the spin) is small (The initial configuration already has the smallest value of energy), the difference between the original configuration and the configurations generated during the Monte Carlo chain in the simulations is nonsignificant. This would lead to the scenario that
the values of $R$ for $T < T_c$ stay close to 1 which is what's been observed in the left panel of fig.~\ref{g01L72L24}. Similarly, at a given temperature, for
a simulation beginning with a uniform-start configuration, 
one of the ground state can be reached by the MC chain. Once
such a ground state configuration is obtained, then subsequent configurations cannot be modified noticeably from that ground state configuration. This kind of argument explains the tunneling between $R = 1$ and $R = 1/\sqrt{2}$ for $T < T_c$ shown in the middle and the right panels of fig.~\ref{g01L72L24}.

\begin{figure}
	\hbox{             
		\includegraphics[width=0.33\textwidth]{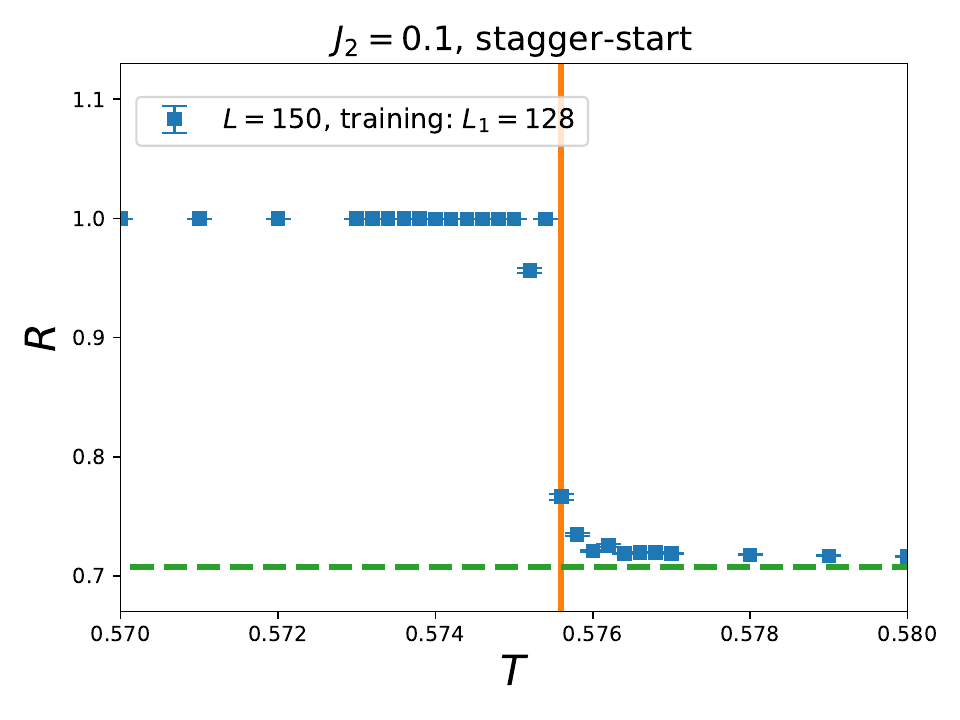}
		\includegraphics[width=0.33\textwidth]{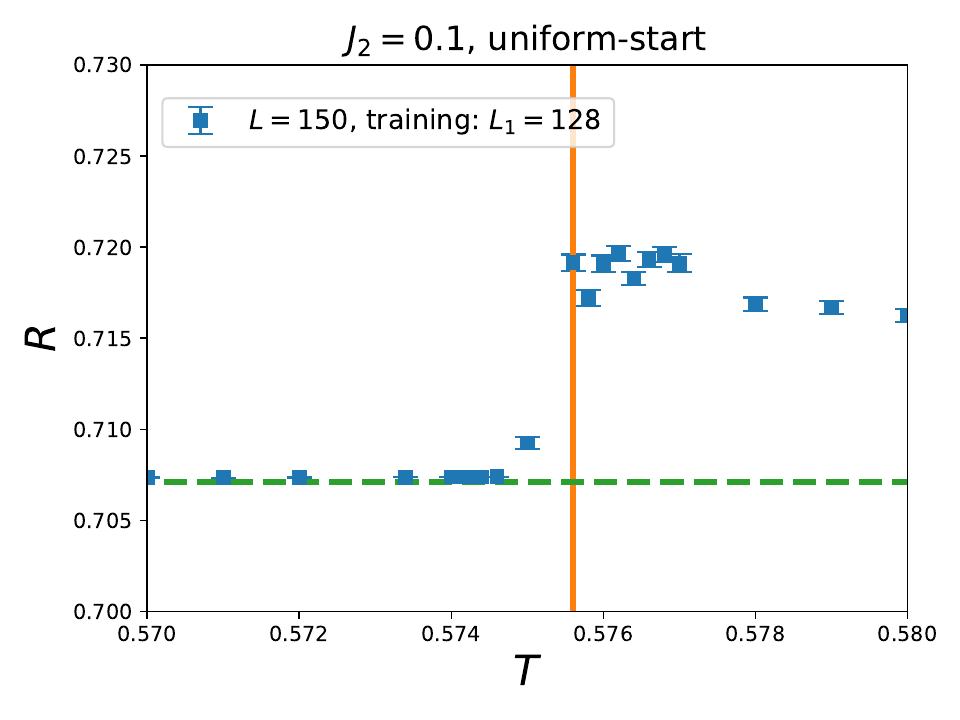}
		\includegraphics[width=0.33\textwidth]{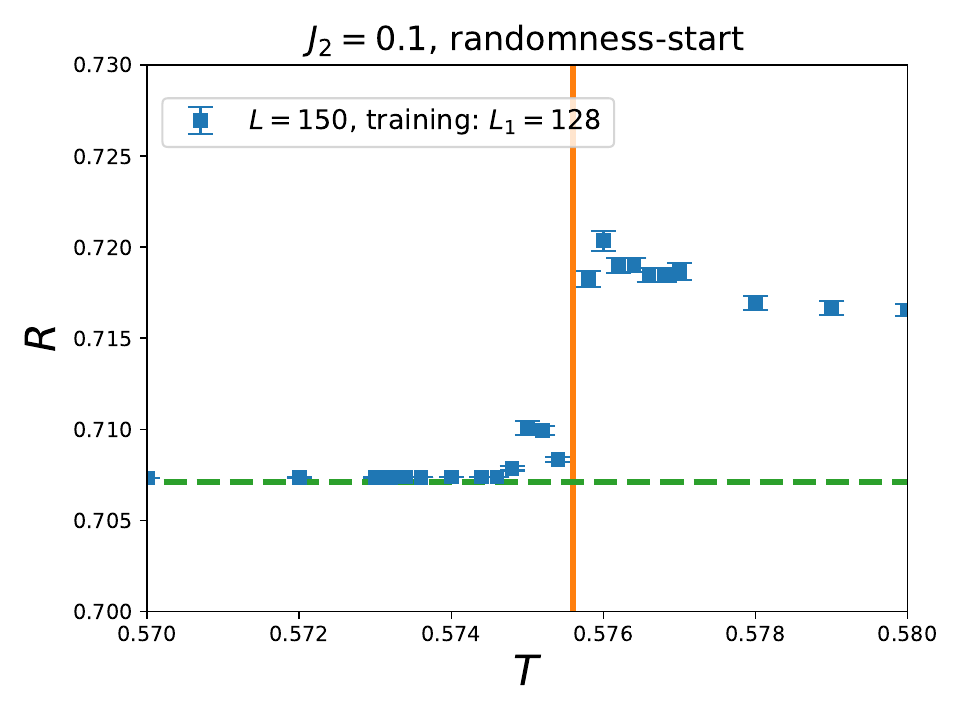}	
	}
	\caption{$R$ as functions of $T$ for the $g=0.1$, $L=150$, and $L_1=128$ close to $T_c$. The vertical solid lines are the expected $T_c$. The dashed horizontal lines are $1/\sqrt{2}$ which is the possible smallest value of $R$.}
	\label{g01L150L128}
\end{figure}

$R$ as functions of $T$ for $L=150$ and $L_1=128$ are shown in
fig.~\ref{g01L150L128}. The message reveals from fig.~\ref{g01L150L128} is the same as that related to fig.~\ref{g01L72L24}. Specifically, the determined value of $T_c$ for $J_2 = 0.1$ agrees well with the known outcome in the literature
and the phase transition is first order.

\subsection{The NN outcomes related to $J_2=0.5$}

\begin{figure}
	\hbox{             
		\includegraphics[width=0.33\textwidth]{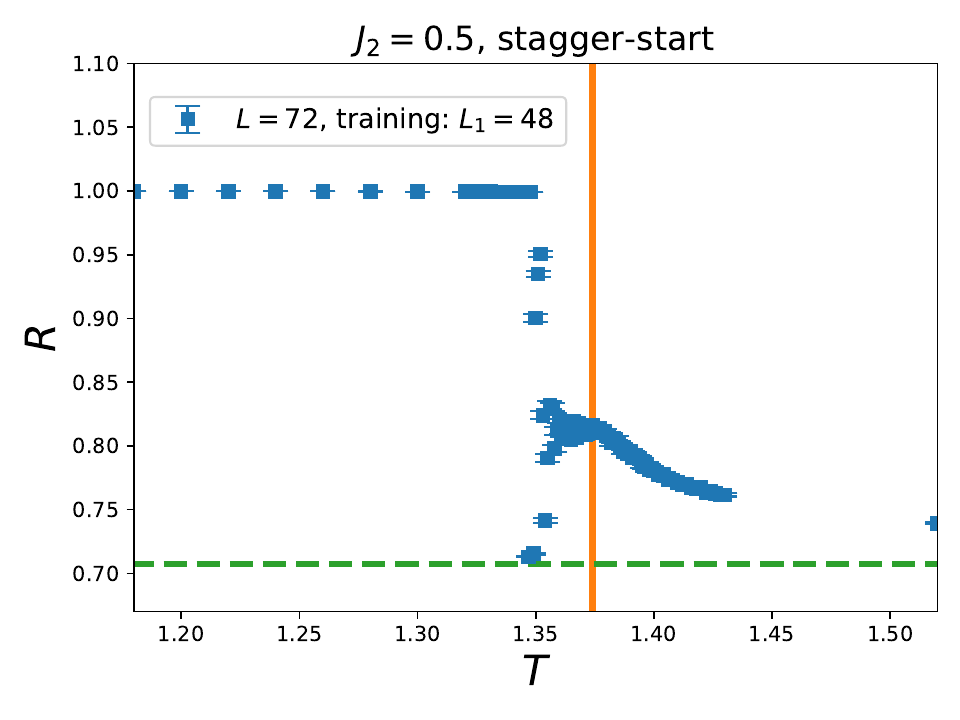}
		\includegraphics[width=0.33\textwidth]{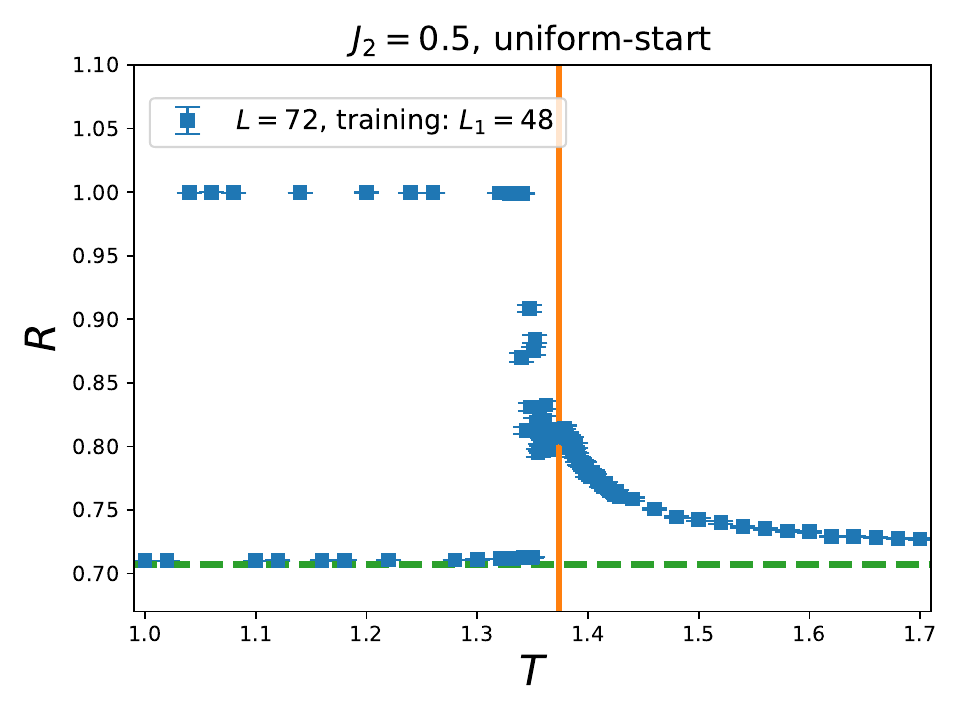}
		\includegraphics[width=0.33\textwidth]{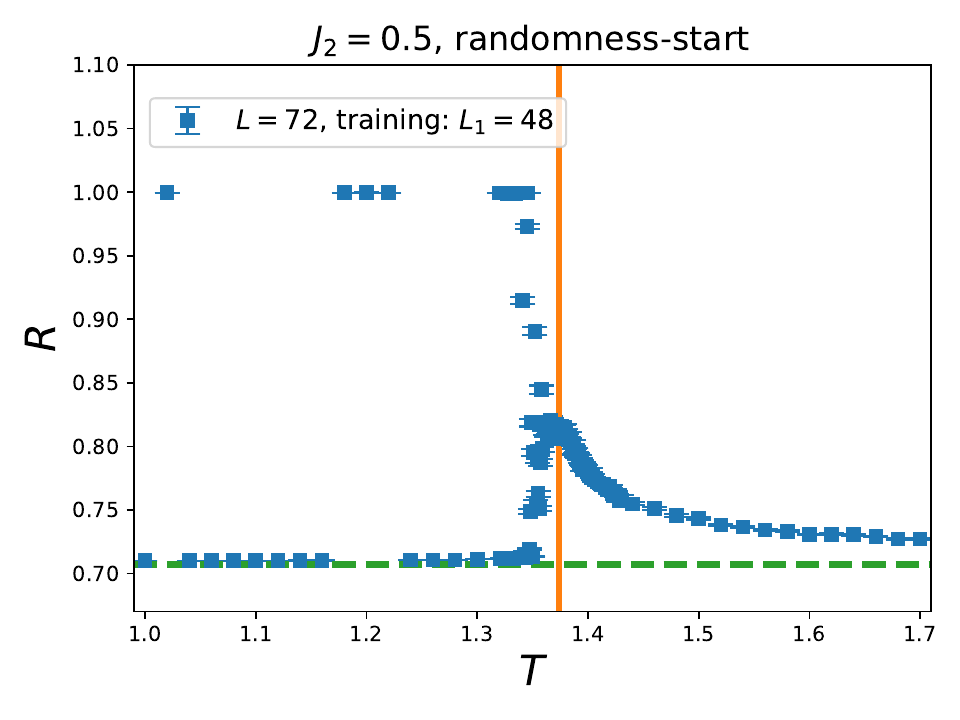}	
	}
	\caption{$R$ as functions of $T$ for the $g=0.5$, $L=72$, and $L_1=48$. The vertical solid lines are the expected $T_c$. The dashed horizontal lines are $1/\sqrt{2}$ which is the possible smallest value of $R$.}
	\label{g05L72L48}
\end{figure}

\begin{figure}
\hbox{             
	\includegraphics[width=0.33\textwidth]{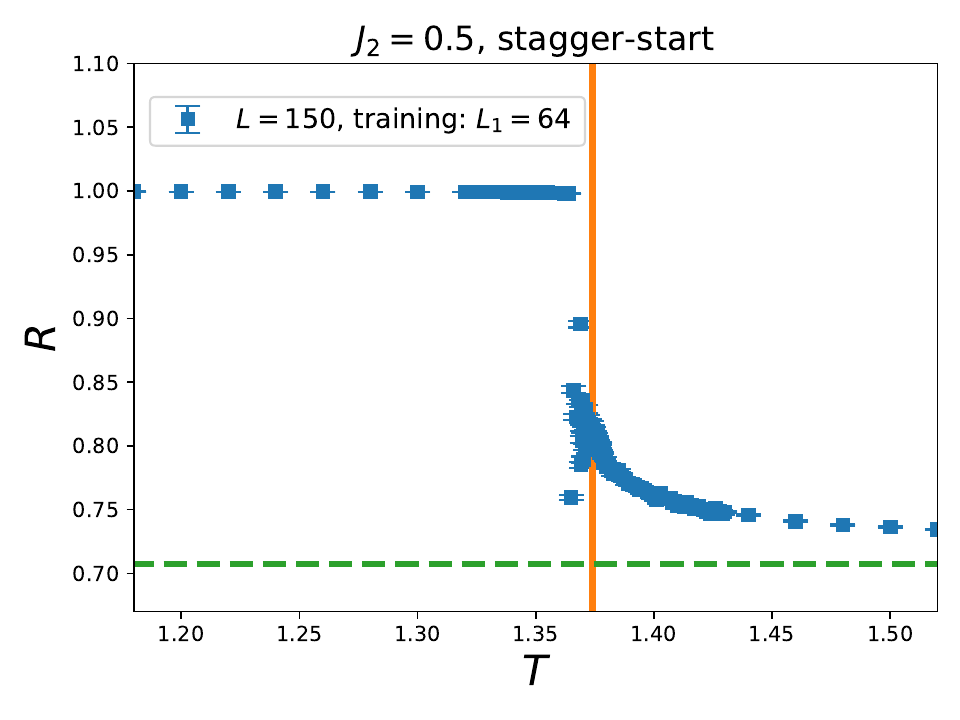}
	\includegraphics[width=0.33\textwidth]{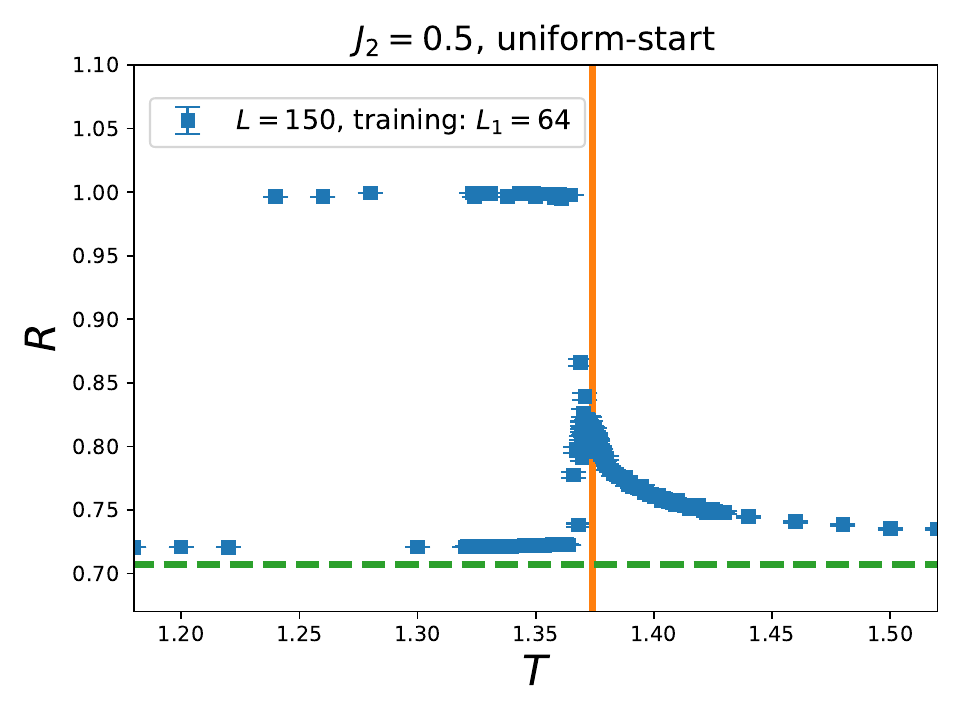}
	\includegraphics[width=0.33\textwidth]{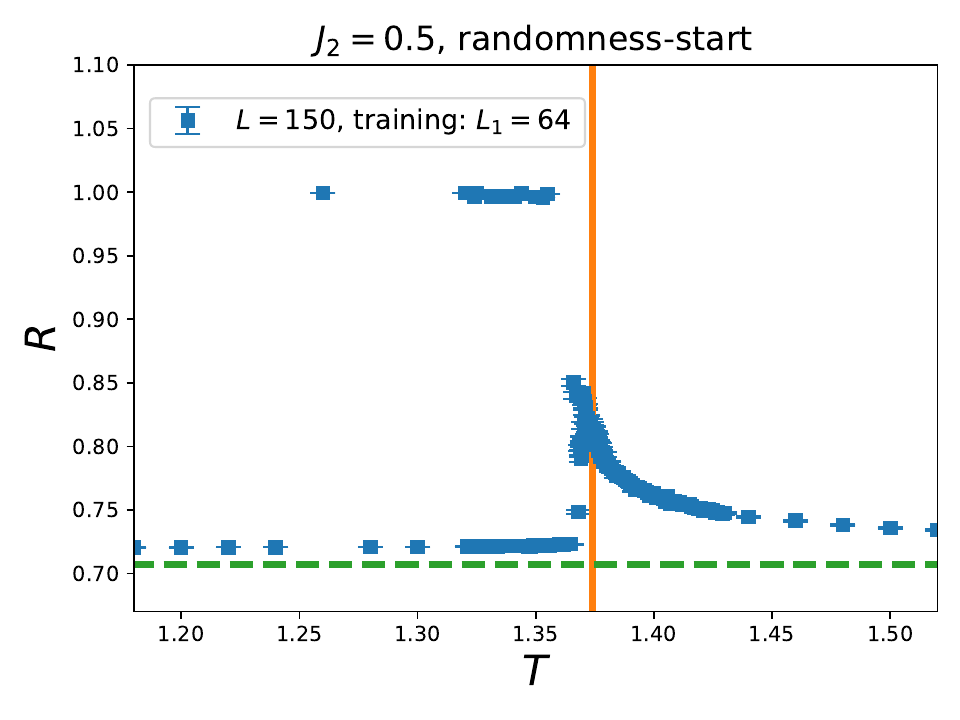}	
}
\caption{$R$ as functions of $T$ for the $g=0.5$, $L=150$, and $L_1=64$. The vertical solid lines are the expected $T_c$. The dashed horizontal lines are $1/\sqrt{2}$ which is the possible smallest value of $R$.}
\label{g05L150L64}
\end{figure}

For $J_2 = 0.5$, the obtained $R$ as functions of $T$ for $L=72$ and $L_1=48$ are shown in
fig.~\ref{g05L72L48}. 
In addition, the obtained $R$ as functions of $T$ for $L=150$ and $L_1=64$ are shown in
fig.~\ref{g05L150L64}. 
The initial configurations used to begin
the Monte Carlo simulations are listed in the titles of the three
panels of figs.~\ref{g05L72L48} and \ref{g05L150L64}. The vertical solid lines in the these figures are the $T_c$ for $g=0.5$ determined in Ref.~\cite{Ras05}. Clearly, discontinuity occurs in $R$ near the expected $T_c$. In particular, for $L=150$ and $L_1=64$, sudden jumps in $R$
is more apparent than that related to $L=72$ and $L_1 = 24$.
This indicates the transition is first order beyond doubt. The results presented in
all the panels of figs.~\ref{g05L72L48} and \ref{g05L150L64} provide convincing evidence
to support the facts that our MLP trained without using any real configurations of the considered models can detect the $T_c$ precisely and can determine the nature of the phase transition correctly.

\begin{figure}
	\hbox{~~~~~~                                                                 
	\includegraphics[width=0.4\textwidth]{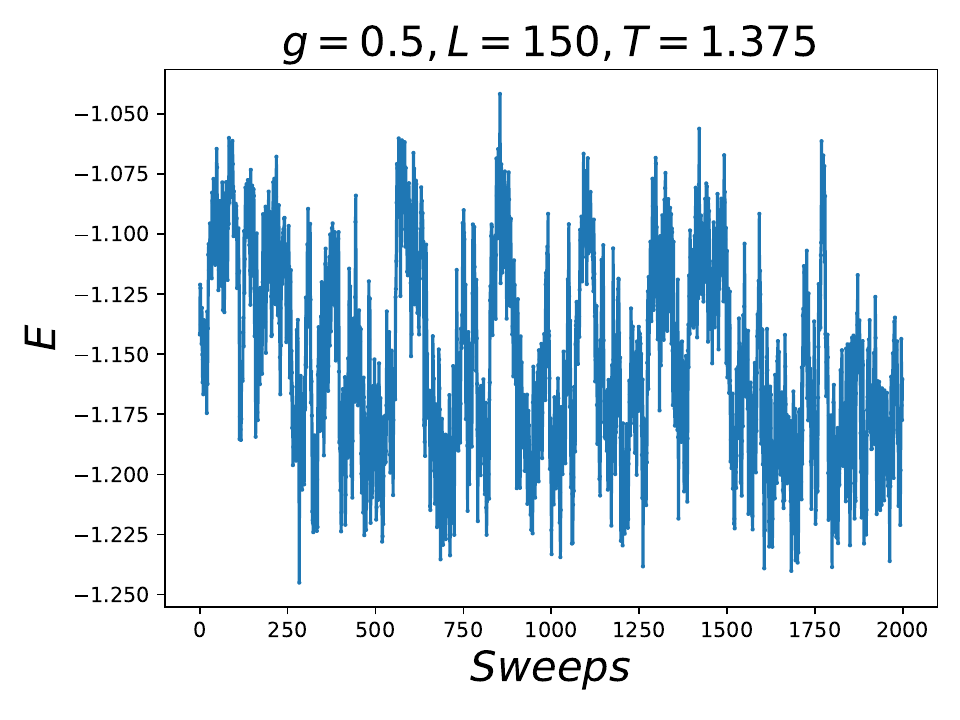}~~~~~~~~~
	\includegraphics[width=0.4\textwidth]{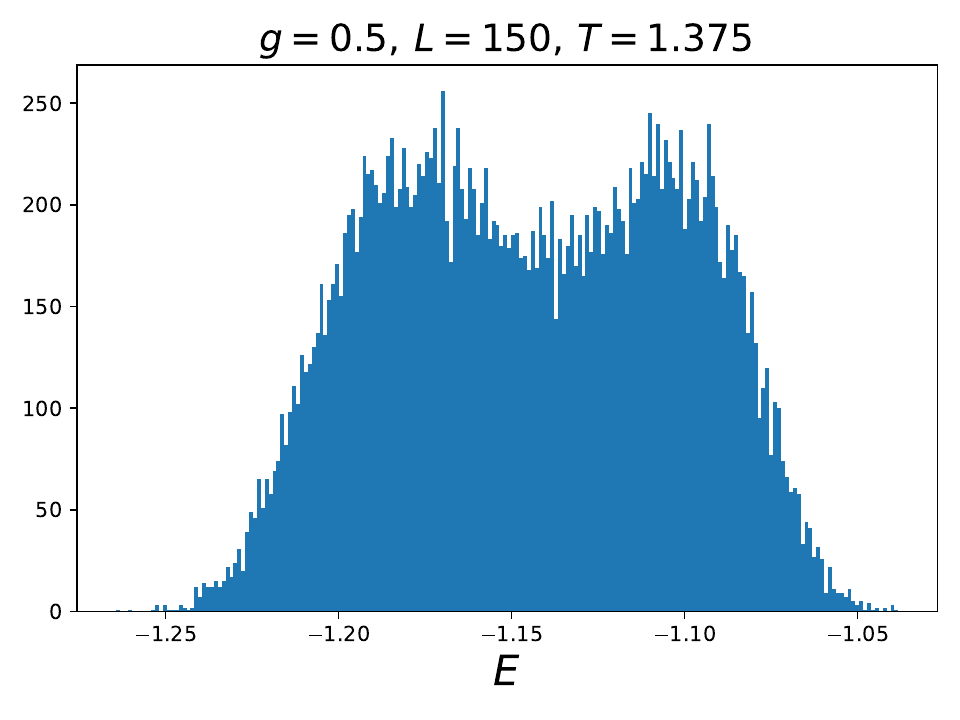}
}
	\caption{The running history (left panel) and histogram (right panel) of $E$ for $J_2=0.5$ with $L=150$.}
	\label{energy}
\end{figure}

It should pointed out that based on the Monte Carlo data of Ref.~\cite{Ras05}, the associated phase transition is a weak first-order phase transition for $J2=0.5$. The first 2000 running history (left panel) and the whole histogram (right panel) of the energy density $E$ for $J_2=0.5$ with $L=150$ are shown in fig.~\ref{energy}. The data are obtained
by performing a measurement once when 100 sweeps are executed. The results shown in
fig.~\ref{energy} are highly similar to the corresponding outcomes given in Ref.~\cite{Ras05}. We would like to emphasize the facts that besides the interpretation of
a signal of weak first-order phase transition, the running history 
of $E$ shown in the left panel of fig.~\ref{energy} can be explained
by the inefficiency of the Monte Carlo algorithm (Hence the correlation of the data is large). In addition, the appearance of two-peak structure
for the histogram of $E$ on a single lattice does not imply non-vanishing latent heat. Indeed, while the data obtained in Ref.~\cite{Ras05}
favor the scenario of a first-order phase transition for $J_2 = 0.5$,
more convincing outcomes are needed to reinforce this claim. One should demonstrate
that either the two-peak structure becomes more noticeable with increasing $L$ or an extremely sharp two-peak structure appears in the histogram of certain relevant quantities.
Finally, it should be pointed out that the specific heat data associated with $g=0.5$ determine in Ref.~\cite{Ras05} do not fulfill the expected scaling law for a first-order phase transition.  

On the other hand, our MLP results shown in fig.~\ref{g05L150L64} ($L=150$ and $L_1=64$) demonstrate a clearer signal of a first-order phase transitions for $g=0.5$ than that of the Monte Carlo outcomes.
This could be considered as one advantage of our unconventional NN approach used in this study. Later, we will show that indeed an extremely sharp two-peak structure shows up in the histogram of $R$ which reinforces the conclusion that our unconventional NN method is exceedingly useful in detecting the considered weak first-order phase transition related to $J_2 = 0.5$.

\subsection{The NN outcomes related to $J_2=1.0$}

\begin{figure}
	\hbox{             
		\includegraphics[width=0.33\textwidth]{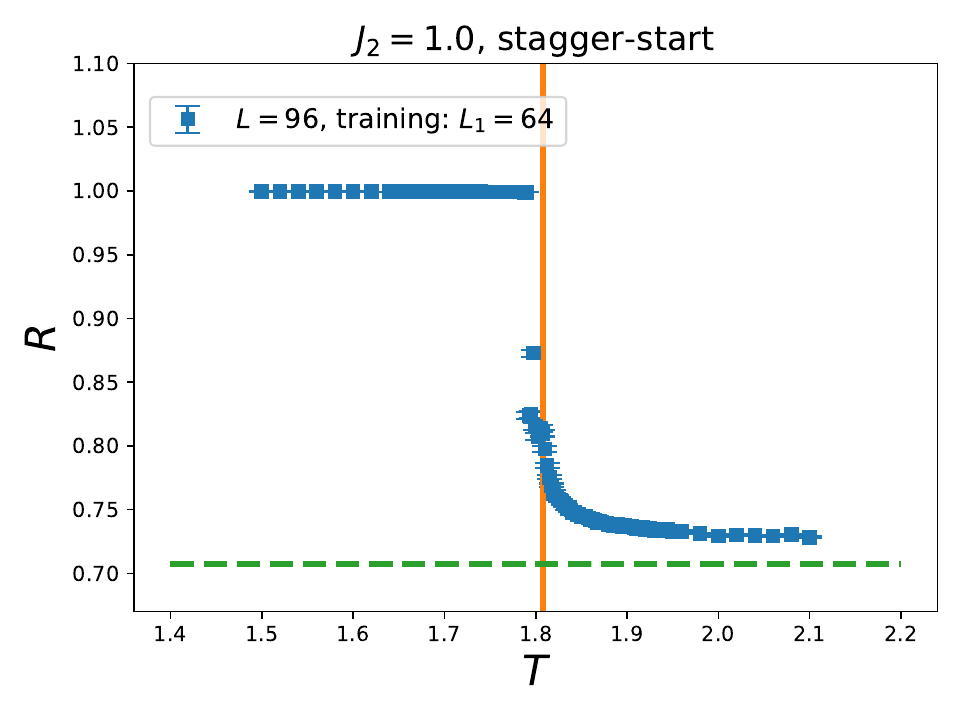}
		\includegraphics[width=0.33\textwidth]{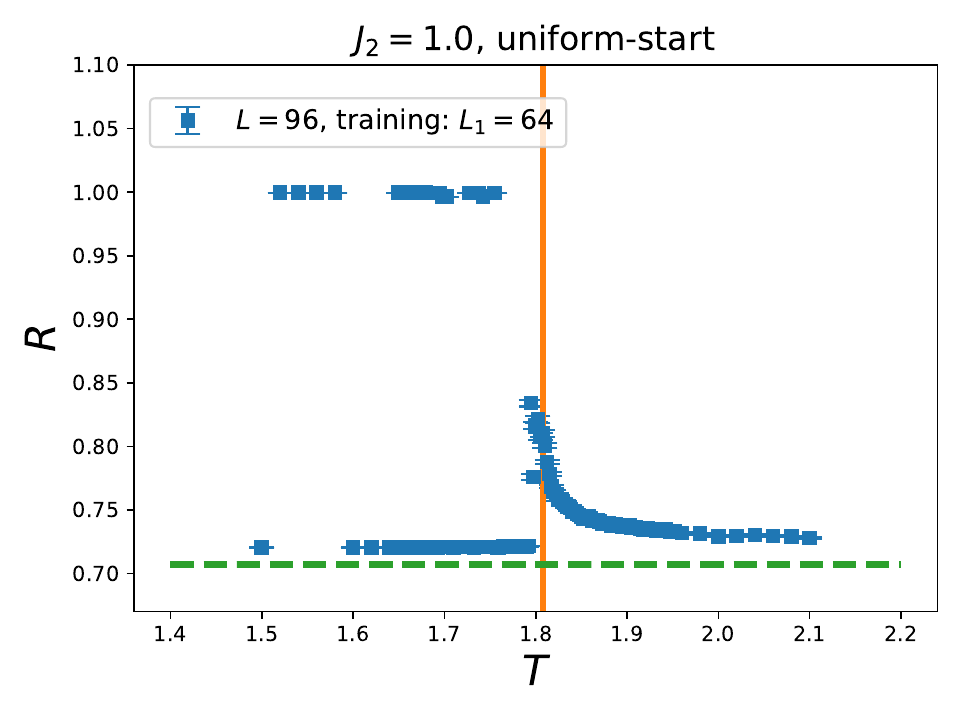}
		\includegraphics[width=0.33\textwidth]{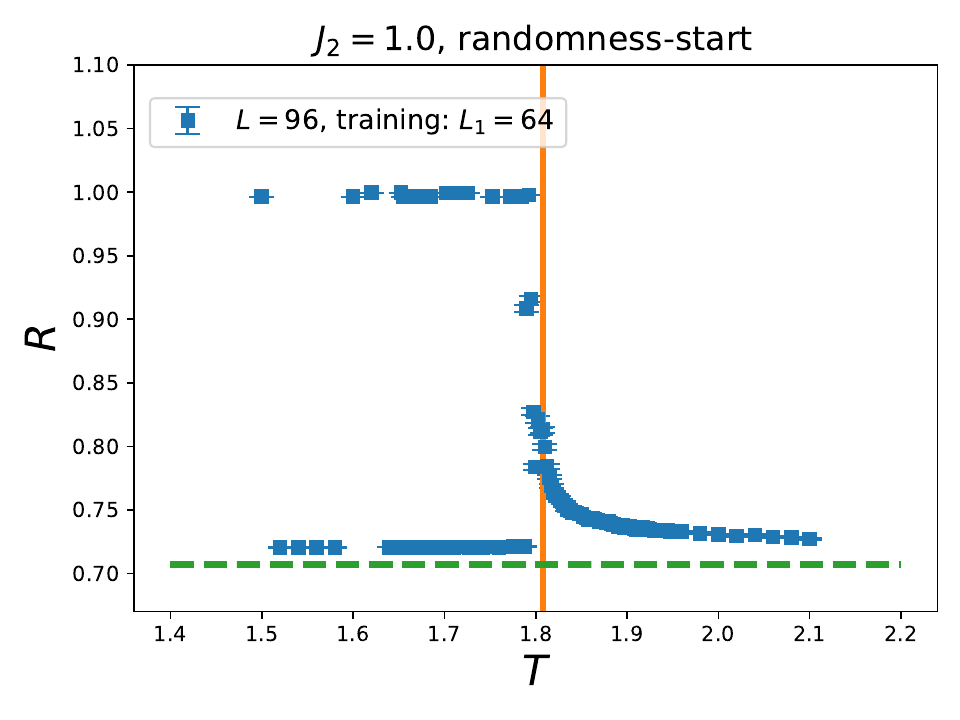}	
	}
	\caption{$R$ as functions of $T$ for the $g=1.0$, $L=96$, and $L_1=64$. The vertical solid lines are the $T_c$. The dashed horizontal lines are $1/\sqrt{2}$ which is the possible smallest value of $R$.}
	\label{g10L96L64}
\end{figure}

For $J_2 = 1.0$, the obtained $R$ as functions of $T$ for $L=96$ and $L_1=64$ are shown in
fig.~\ref{g10L96L64}. In addition, the initial configurations used to begin
the Monte Carlo simulations are listed in the titles of the three
panels of fig.~\ref{g10L96L64}. The vertical solid lines in the fig.~\ref{g10L96L64} are the $T_c$ for $g=1.0$ determined in Ref.~\cite{Ras05}. Clearly, sudden jumps in $R$ take place at a value of $T$ that agrees well with the expected $T_c$. This indicates the transition is first order beyond any doubt. The results presented in
all three panels of fig.~\ref{g10L96L64} provide convincing evidence
to support the facts that our MLP trained without using any real configurations of the considered models can detect the $T_c$ precisely and can determine the nature of the phase transition correctly.

\begin{figure}
	\hbox{             
		\includegraphics[width=0.33\textwidth]{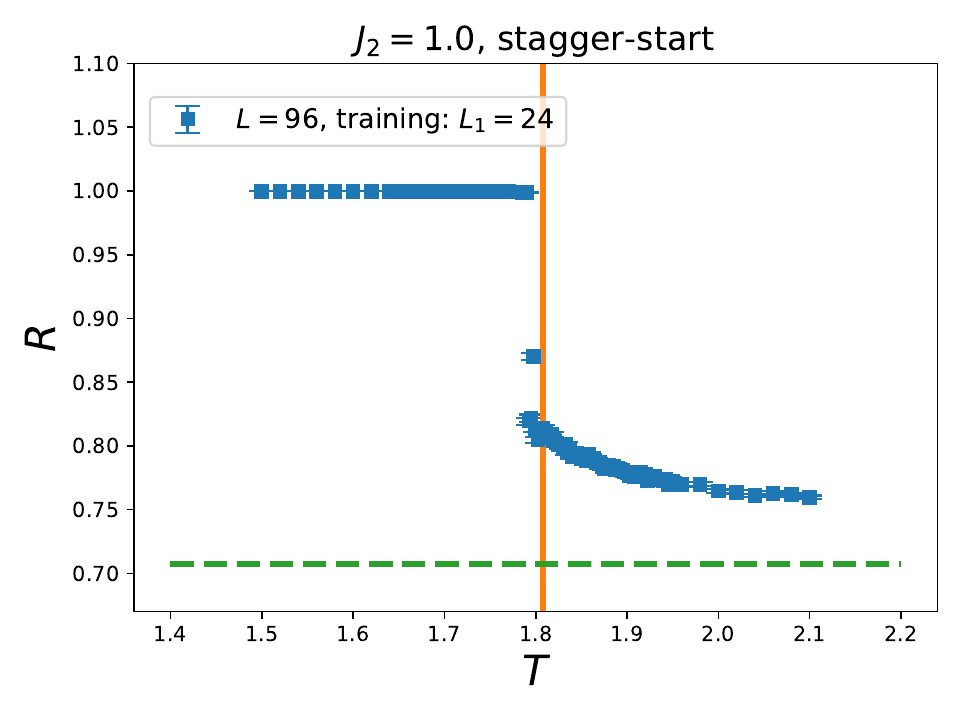}
		\includegraphics[width=0.33\textwidth]{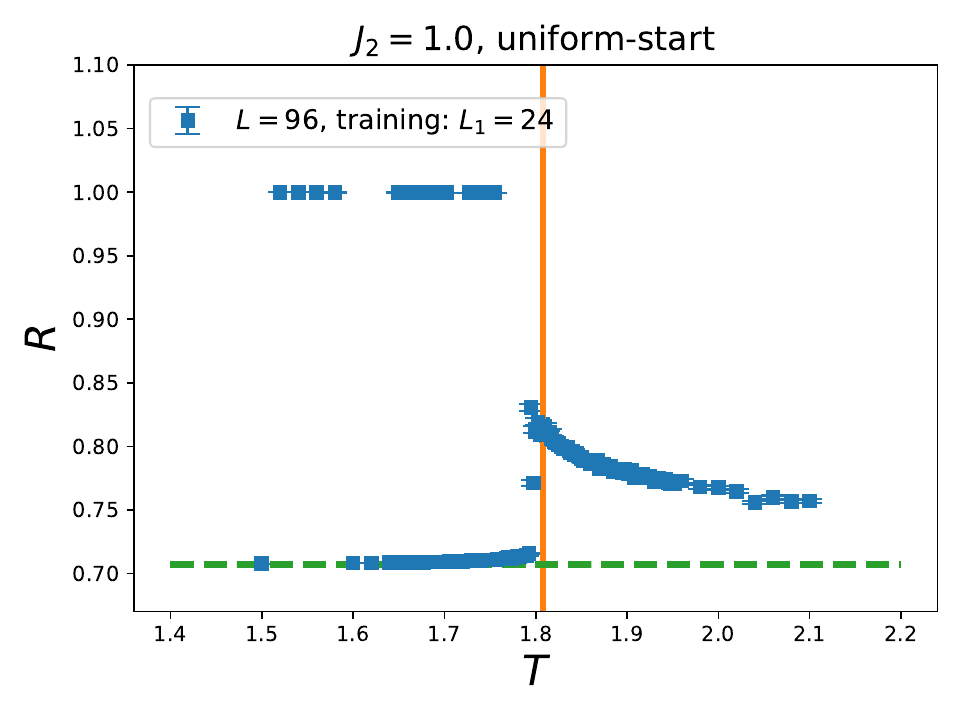}
		\includegraphics[width=0.33\textwidth]{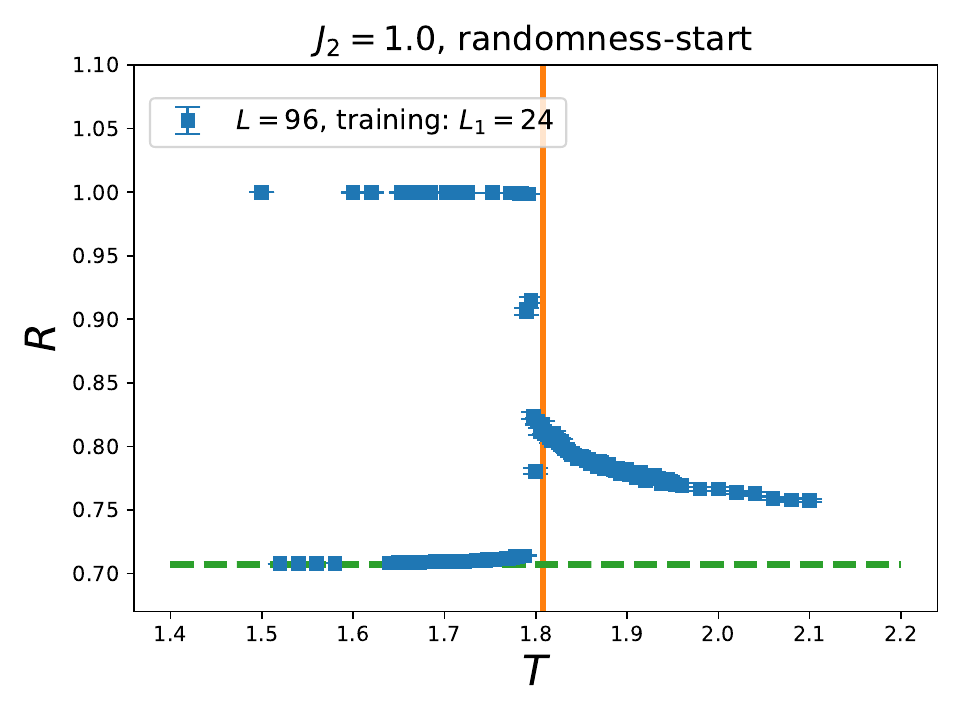}	
	}
	\caption{$R$ as functions of $T$ for the $g=1.0$, $L=96$, and $L_1=24$. The vertical solid lines are the expected $T_c$. The dashed horizontal lines are $1/\sqrt{2}$ which is the possible smallest value of $R$.}
	\label{g10L96L24}
\end{figure}

\begin{figure}
	\hbox{             
		\includegraphics[width=0.33\textwidth]{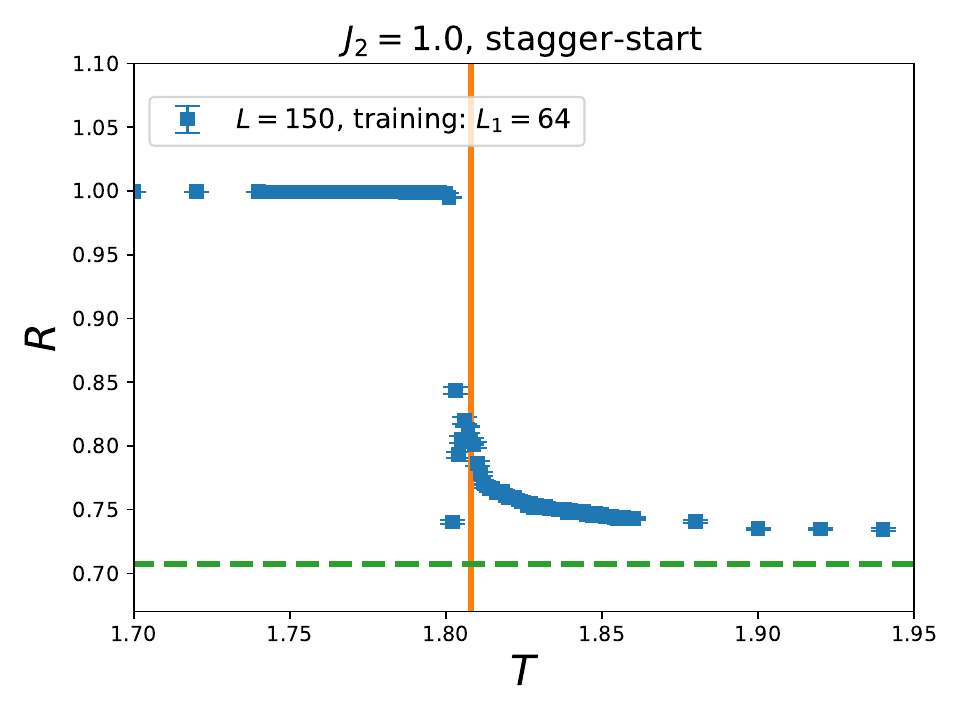}
		\includegraphics[width=0.33\textwidth]{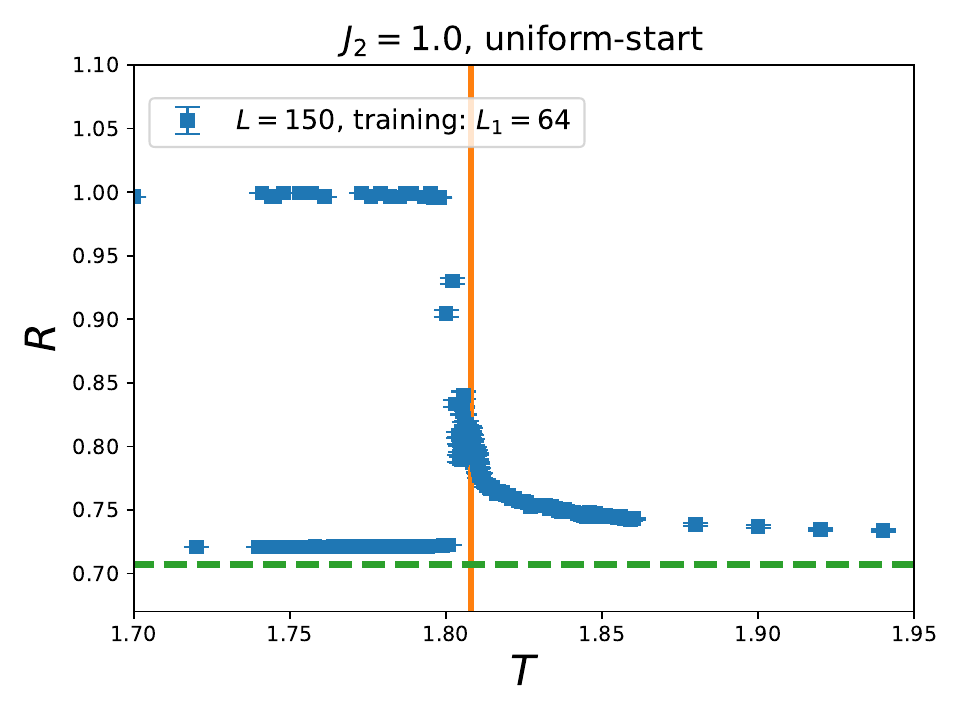}
		\includegraphics[width=0.33\textwidth]{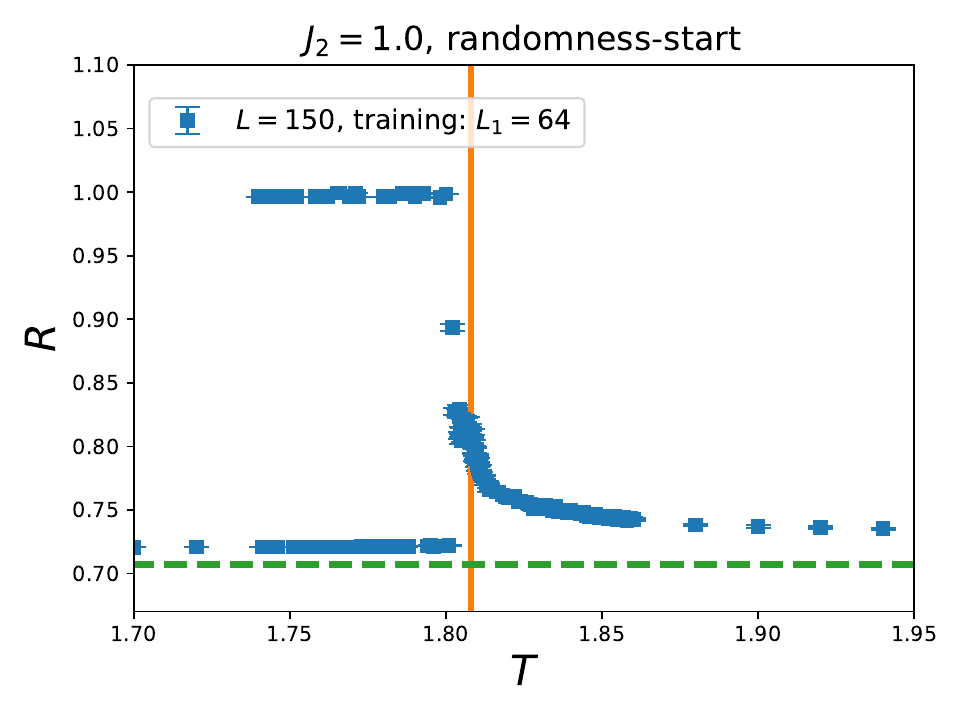}	
	}
	\caption{$R$ as functions of $T$ for the $J_2=1.0$, $L=150$, and $L_1=64$. The vertical solid lines are the expect $T_c$. The dashed horizontal lines are $1/\sqrt{2}$ which is the possible smallest value of $R$.}
	\label{g10L150L64}
\end{figure}

If $L=96$ ($L=150$) and $L_1=24$ ($L_1 = 64$) are considered, the conclusion remain 
the same. In particular, the phase transition for $J_2 = 1.0$
is first order beyond doubt since there is clear discontinuity near
the expected $T_c$, see fig.~\ref{g10L96L24} (fig.~\ref{g10L150L64}).
The outcomes shown in figs.~\ref{g10L96L64}, \ref{g10L96L24},
and \ref{g10L150L64} also indicate that the use of various $L$ and
$L_1$ does not have noticeable impact on the determinations of the $T_c$ and the nature of the phase transition.

\section{Discussions and Conclusions}

In this study, we apply the NN technique to investigate the phase transitions of the 2D antiferromagnetic Ising model with nearest $J_1 = 1$ and next-to-nearest $J_2 > 0$ interactions on the triangular lattice. In 
particular, we focus on determining the nature of phase transitions for $J_2 = 0.1, 0.5$ and 1.0. The NN outcomes lead to the conclusions that the
these phase transitions are first order. Moreover, the first-order
phase transition for $J_2=0.5$ is a weak one.

We would like to emphasize the fact that 
we employ a simple MLP having a hidden layer with only two neurons (or other number of neurons).
In addition, the training set are not from the real spin configurations, but are made artificially. Particularly,
in the training set, there are merely two types of configurations having their spins being ordered in a staggered pattern.

Previously, a similar training strategy of using artificially made
configurations as the training set appears in Ref.~\cite{Tan20.1}. In particular,
the corresponding spins are arranged in a ferromagnetic pattern. It has been demonstrated that this training setup can accurately compute the critical points of many different models such as the 2D ferromagnetic $q$-state Potts models on the square lattice (for $q=2,3,4,6,8,9,$ and 10) \cite{Tse241}, the 2D classical and generalized $XY$ models on the square lattice \cite{Tse22}, the 3D classical $O(3)$ model on a cubic lattice \cite{Tan20.1}, and so on.

The conventional NN calculations use real whole configurations as the training sets and the testing sets. These are files of huge sizes. As a result, it takes a lot of time to train the NN and the required storage space for these files are anticipated to be large. Besides, when a different linear system size $L$ or another model is considered, one typically needs to train a new NN for that particular situation. 
The unconventional training approach of Ref.~\cite{Tan20.1} not only speeds up the NN computations by a factor of at least several hundred, but is also applicable to many models, namely it is universal. It is anticipated that the new training strategy employed in this study has the same advantages. To demonstrate this, investigating 
the phase transitions of other models with untypical phase transitions or nontrivial ground state configurations, such as the 2D 3-state
and 4-state antiferromagnetic Potts models on the square lattice
and the 2D frustrated $J_1$-$J_2$ Ising model on the square or honeycomb lattices, should be conducted using the built MLP in this study.

We would like to emphasize the fact that if one embeds the triangular lattice in an area with different geometry than the one used here, then the arrangement of spins with values 1 and -1 of a ground state configurations may
seem to be varied from the ones shown here, see Ref.~\cite{Mut24}
for one example. 

\begin{figure}
	\hbox{		
		\includegraphics[width=0.33\textwidth]{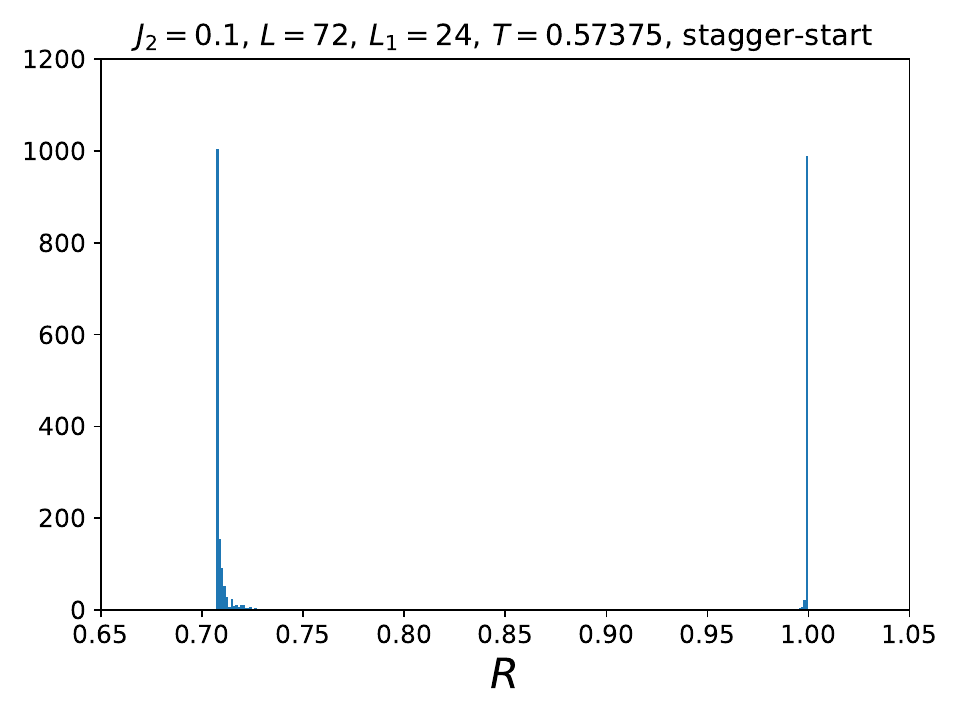}
			\includegraphics[width=0.33\textwidth]{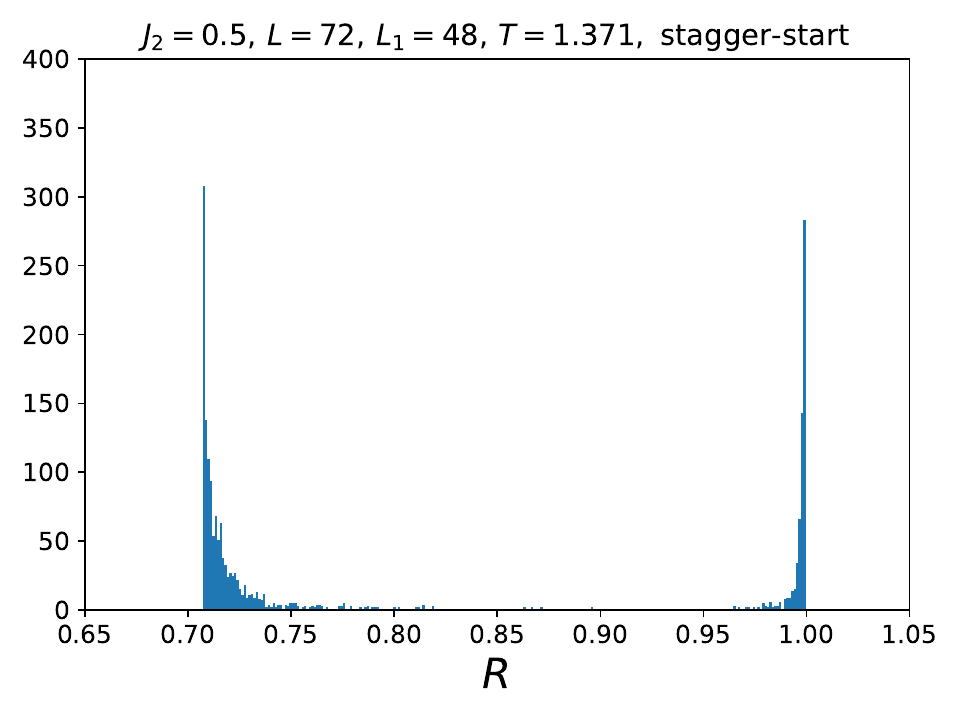}
			\includegraphics[width=0.33\textwidth]{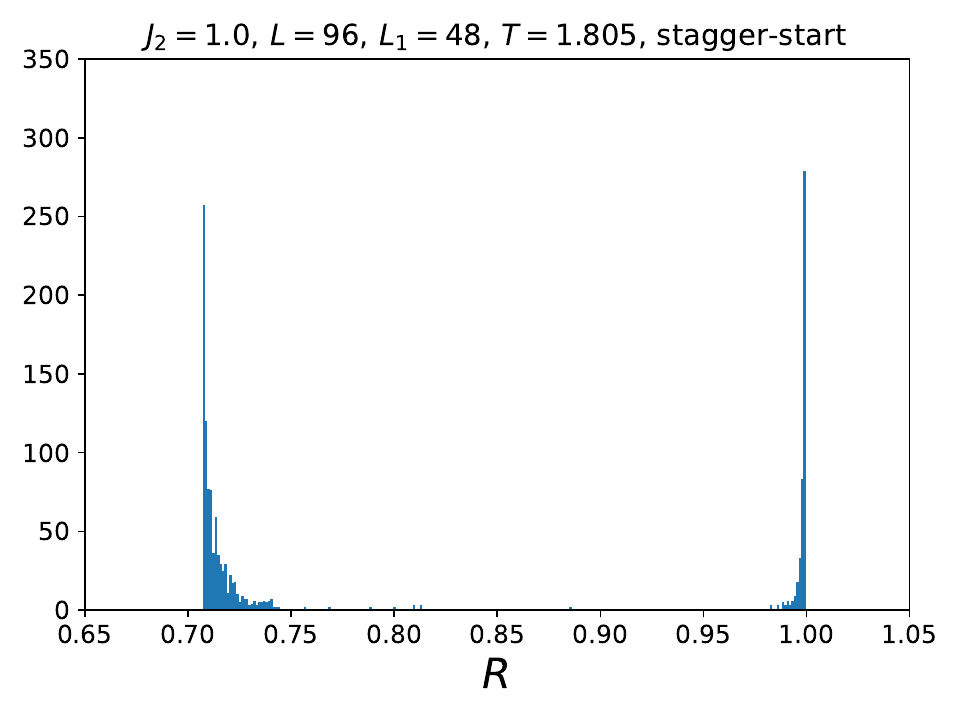}
		}
	\caption{(Left) The histogram of $R$ for $J_2=0.1$, $L=72$, $L_1=24$ and $T=0.57375$. (Middle) The histogram of $R$ for $J_2 = 0.5$, $L=72$, $L_1 = 48$ and $T=1.731$. (Right) The histogram of $R$ for $J_2 = 1.0$, $L=96$, $L_1 = 48$ and $T=1.805$.
	The data are obtained with stagger-start.}
	\label{histogram}
\end{figure}

One may argue that the scenario of sudden jump(s) in $R$ near a particular value of $T$ when $R$ is considered as a function of $T$ does not provide sufficient evidence to support the claims that these studied phase transition are first order. We would like to point that by 
investigating the running histories or the histograms of $R$, typical behaviors (or the conventional signal) of a first-order phase transition appear. For instance, the histogram of $R$ for $J_2 = 0.1$, $L=72$, and $T=0.57375$ is shown in the left panel of fig.~\ref{histogram}. Clearly, almost all the $R$ take the values of either 1 or $1/\sqrt{2}$. This is exactly a feature of a first-order transition. Moreover, the histogram of $R$ for $J_2 = 0.5$, $L=72$, $L_1 = 48$, and $T=1.731$ is shown in the middle panel of fig.~\ref{histogram}. The signal of the first-order phase transition shown in the middle panel of fig.~\ref{histogram} for $J_2 = 0.5$ is much stronger than that of the right panel of fig.~\ref{energy}. This demonstrates the powerfulness of our unconventional NN approach for uncovering the nature of a phase transition. Finally, the same scenario is observed for $g=1.0$, see the right panel of fig.~\ref{histogram}.
In conclusion, $R$ can really serve as an extremely useful criterion to determine whether the nature of a phase transition is first order or second order.

As pointed out previously, the same MLP model (fig.~\ref{nn}) trained with two artificially made (ferromagnet-like) configurations successfully computes the critical points and critical exponents of a few physical models with reasonably high precision \cite{Tan20.1,Tse22,Tse23,Tse241,Tse242,Jia24,Tse25}. 
The $L_2(1)$ regularization with the input parameter being 1 is also used for these studies. This setup of considering kernel $L_2(1)$ regularization works well for both the training strategies employed in this investigation and the mentioned studies. 
For other models with distinct training strategies, the best choice of $d$ for $L_2(d)$ may be different from ours.  
 
Finally, it is also important to emphasize the fact that if one employs the traditional method like Monte Carlo simulations to study the phase
transitions of the models considered here, then one must know the ground state configurations so that the order parameter magnetization $m$ can be constructed and used to investigate the associated critical behavior. For instance, in Ref.~\cite{Ras05},
the quantity magnetization $m$ is built as $m = \sqrt{\left(m_1^2 + m_2^2\right)/2} $, where $m_1$ and $m_2$ are given as 
\begin{equation}	
m_i = \frac{2}{L^2}\sum_{j \in i} \sigma_j,		
\end{equation}
and $m_i$ is the sublattice magnetization ($i = 1,2$).
In other words, the knowledge of the how the configurations are ordered is required to carry out the investigation. This knowledge is in principle not needed when the study is conducted with the NN approach. 

Of course, for the MC calculations, one can use the quantities 
energy density and the associated specific heat to investigate some targeted critical behaviors. Still, no information of ground states, particularly their ordered pattern, can be gained. 
In this study, the result that $R = 1$ at some low temperatures provides strong evidence to support the fact that configurations with their spins being arranged in a staggered pattern is one of the ground states. Interestingly, a close look of the top left panel of fig.~\ref{initial} indicates that a stripe pattern appears if one moves along the axes parallel to the boundaries of parallelogram.  
This observation with the outcome that $R \sim 1/\sqrt{2}$ (This meas the corresponding configuration is in strong contrast to those in the training set) at some 
low temperatures suggest that stripe configurations, with spins taking the same value along rows (or columns) in an alternative manner, are likely the pattern of ground states as well.
To prove this from the NN results, a more systematic investigation is required and is left as a future study. Finally, the stripe configurations described in this paragraph are exactly the ground states mentioned in Refs.~\cite{Ras05,Mut24}.

Finally, let us summarize the features (and advantages) of our NN approach
considered in this study.

\begin{enumerate}
	\item{Two-type of artificially made configurations are used as the training set. In particular, the spins with their value being 1 and -1 are arranged in a staggered pattern. }
		\item{No real physical quantities, such as the spin configurations, are employed to train the built MLP. }
		\item{The obtained MLP trained without any physical inputs from the considered model computes the $T_c$ accurately and determines the nature of the studied phase transitions correctly.}
		\item{The training can be conducted only once with a given fixed $L_1$, and the resulting MLP is capable of detecting the $T_c$ ($T_c(L)$) precisely for many values of $J_2$ and various $L$.}
		\item{The features mentioned above speed up our NN computation by a factor of
		at least several hundred times when compared with the conventional procedures typicaly considered in the literature. Apart from this, the required storage capacity is a few times smaller than that usually needed for a standard NN calculation as well.  }			
		\item{For the considered model, the signal of first-order phase transitions, such as the appearance of two-peak structure in the histograms of certain observables, obtained from our NN calculations are much stronger than that determined from the Monte Carlo simulations.}
	        \item{The obtained MLP is likely universal, namely it can directly adopted to study the critical phenomena of other models without being retrained.
                To confirm this, further investigations should be carried out.}		
\end{enumerate}

\section*{Funding}\vskip-0.3cm
Partial support from National Science and Technology Council (NSTC) of
Taiwan is acknowledged (Grant numbers: NSTC 113-2112-M-003-014- and NSTC 114-2112-M-003-004-).

\section*{Acknowledgment}
The MLP used in this study are implemented using 
the library Keras of TensorFlow \cite{tens}.

\section*{Conflict of Interest}
The authors declare no conflict of interest.

\section*{Data Availability Statement}
Data are available from the corresponding author
on reasonable request.

\end{document}